\documentclass[aps,prb,twocolumn,groupedaddress,showpacs]{revtex4-1}

\usepackage{graphicx}
\graphicspath{ {/} }

\newcommand{\mytilde}{\raise.17ex\hbox{$\scriptstyle\mathtt{\sim}$}}

\begin{document}


\title{Effect of interfacial intermixing on the Dzyaloshinskii-Moriya interaction in Pt/Co/Pt}


\author{Adam W. J. Wells}
\email[]{pyaww@leeds.ac.uk}
\author{Philippa M. Shepley}
\author{Christopher H. Marrows}
\author{Thomas A. Moore}
\affiliation{School of Physics and Astronomy, University of Leeds, Leeds LS2 9JT, United Kingdom}


\date{\today}

\begin{abstract}
We study the effect of sputter-deposition conditions, namely substrate temperature and chamber base pressure, upon the interface quality of epitaxial Pt/Co/Pt thin films with perpendicular magnetic anisotropy. Here we define interface quality to be the inverse of the sum in quadrature of roughness and intermixing. We find that samples with the top Co/Pt layers grown at 250$^\circ$C exhibit a local maximum in roughness-intermixing and that the interface quality is better for lower or higher deposition temperatures, up to 400$^\circ$C, above which the interface quality degrades. Imaging the expansion of magnetic domains in an in-plane field using wide-field Kerr microscopy, we determine the interfacial Dzyaloshinskii-Moriya interaction (DMI) in films in the deposition temperature range 100$^\circ$C to 300$^\circ$C. The net DMI is linked to the difference in top and bottom Co interface qualities; the net DMI increases as the difference between top and bottom Co interface quality increases. Furthermore, for sufficiently low base pressures, the net DMI increases linearly with the deposition temperature, indicating that fine-tuning of the DMI may be achieved via the deposition conditions.
\end{abstract}


\maketitle

\section{\label{Introduction}Introduction}

The interfacial Dzyaloshinskii-Moriya interaction (DMI) is a key ingredient in determining the equilibrium domain wall (DW) spin structure in thin magnetic films with perpendicular anisotropy and structural inversion asymmetry, such as Pt/Co/AlOx \cite{Pizzini14, Belmeguenai15, Benitez15}, Pt/Co/Ir \cite{Hrabec14, Yang16}, Pt/[Co/Ni] \cite{Koyama11, Ryu13, Chen13} etc.  The DW spin structure in turn determines how the DW responds to a driving force.  In the presence of DMI, bubble domains expand asymmetrically in simultaneously applied in-plane and out-of-plane fields\cite{Je13, Hrabec14, Lavrijsen15, Vanatka15, Ummelen15}, which enables evaluation of the DMI and of the DW spin structure.  For sufficiently large DMI, N\'{e}el walls are stable\cite{Emori13, Thiaville12} and have been found to move at several 100 m/s under spin-orbit torque\cite{Vanatka15, Thiaville12, Ryu13}.  Beyond that, skyrmions may be stabilized\cite{Sampaio13, Jiang15, Moreau-Luchaire16, Woo16, Boulle16} and could have a huge impact on magnetic memory\cite{Kang16, Parkin08} and logic devices\cite{Zhang15, Allwood05}.

Since DMI originates at the interfaces of a thin magnetic film\cite{Fert90}, contributions from the top and bottom interface must differ in magnitude or sign to effect a net DMI. Previously it has been shown that even nominally symmetric Pt/Co/Pt possesses DMI\cite{Je13, Hrabec14, Lavrijsen15, Moon15, Petit15}.  Bubble domains in room-temperature sputtered Pt/Co/Pt on a silicon substrate expand asymmetrically in an applied in-plane field\cite{Hrabec14}, indicating a net DMI. However, if Pt/Co/Pt is grown epitaxially on sapphire, the domain expansion can be symmetric\cite{Hrabec14}, indicating that there is no net DMI in this case. This highlights the importance of structure and the relative interface morphology of upper and lower Co interfaces in determining the DMI.

Here we adjust the interface morphology of the upper Co interface relative to the lower Co interface by controlling the substrate temperature during deposition.  We find that Co-Pt intermixing increases in the temperature range 100-250$^\circ$C, and correlates with an increased magnitude of the DMI field, which is more pronounced at lower base pressures. Our results show that the interfacial DMI depends very sensitively on the ferromagnet/heavy metal interface morphology and thus on film deposition conditions such as substrate temperature as well as chamber pressure.

\section{\label{Structural}Sample deposition}

The samples investigated here are composed of a Pt(3nm)/Co(0.7nm)/Pt(1nm) epitaxial stack deposited by DC-magnetron sputtering onto a C-plane sapphire (Al$_{2}$O$_{3}$) (0001) substrate previously annealed at 700$^\circ$C for 4 hours. The Pt seed layer was sputtered with the substrate held at 550$^\circ$C for optimum smoothness\cite{Mihai13}. Measurements of the residual resistivity ratio (RRR) and the full width half maximum (FWHM) of X-ray diffraction rocking curves in previous work\cite{Mihai13} showed that sputtering onto a C-plane sapphire substrate held at a temperature between 450-550$^\circ$C yielded high quality crystalline Pt films. The Co and top Pt layer were then sputtered with the substrate held at a temperature in the range 50-500$^\circ$C to aid epitaxy, focusing between 100$^\circ$C and 300$^\circ$C where a high degree of crystallographic ordering was found previously\cite{Mihai13}. Films were grown with the base pressure in the range 0.6-3.5$\times$10$^{-7}$ Torr, measured immediately prior to deposition of the seed layer. The Ar pressure needed to obtain stable plasmas ranged from 2.6-3.9$\times$10$^{-3}$ Torr with the lower base pressures requiring a higher Ar working pressure.

\begin{figure}[t]
\includegraphics[scale=0.4]{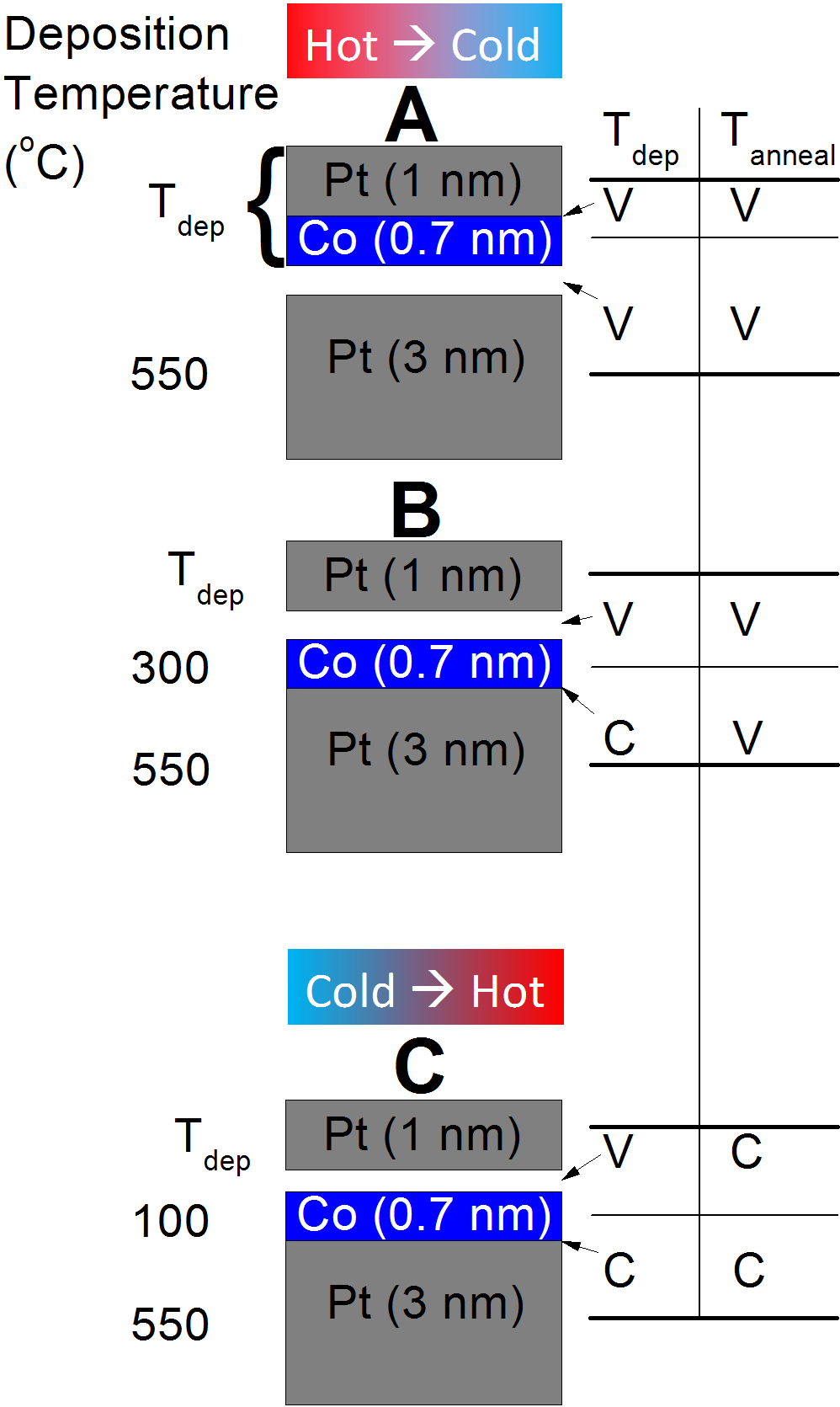}%
\caption{\label{SampleSetTypesGraph} Definition of sample sets A, B and C according to which deposition conditions vary (V) or stay constant (C) between samples, with respect to the top and bottom Co interfaces in a Pt(3nm)/Co(0.7nm)/Pt(1nm) stack. The deposition conditions indicated are the deposition temperature, $T_\mathrm{dep}$, and the annealing temperature, $T_\mathrm{anneal}$. $T_\mathrm{anneal}$ is defined as the highest temperature experienced by the completed stack.}
\end{figure}

In a typical deposition run, a set of samples was grown with the Co and top Pt layer of the first sample deposited at the highest temperature (e.g. 300$^\circ$C) and of the last sample deposited at the lowest temperature (e.g. 100$^\circ$C).  This typical set, which we term type A, forms the basis of this study: detailed structural characterisation of these films is reported in section \ref{Structural} which permits conclusions to be drawn about the dependence of the DMI on the interface morphology, reported in section \ref{DMI}.  Other sets of samples were grown in order to determine the contribution of each interface to the perpendicular anisotropy, reported in section \ref{Magnetic}. To keep the bottom Pt/Co interface morphology constant and modify the top Co/Pt interface only, a set was grown with the Co layer deposited at a fixed temperature and the deposition temperature of the top Pt layer varied (type B). A third set was grown akin to type B but with the order of deposition reversed such that the lowest temperature samples were grown first and the highest temperature samples grown last (type C). It was deduced that the anisotropy originates at both top and bottom Co interfaces, and that it is unaffected by differences in interface quality in the type A films, allowing conclusions about the DMI to be drawn for a set of films where the interface quality varies but the anisotropy remains constant.

As all samples in the growth chamber are heated simultaneously, the maximum temperature a sample experiences may be different from its deposition temperature ($T_\mathrm{dep}$), and this we term the annealing temperature ($T_\mathrm{anneal}$).  For example, in films of type C, samples grown at 100$^\circ$C at the beginning of the run have effectively been annealed at 300$^\circ$C by the end of the run, whereas for films of types A and B, $T_\mathrm{anneal}$ is always less than $T_\mathrm{dep}$. Fig. \ref{SampleSetTypesGraph} organises information \emph{by interface} on the deposition temperature and annealing temperature of each set of films. We also consider the chamber base pressure ($P_\mathrm{base}$) to be a deposition condition: while the \emph{recorded} base pressure is that obtained immediately prior to deposition of the Pt seed layer, $P_\mathrm{base}$ in fact decreases gradually during a deposition run as a result of outgassing, and for the formation of a given interface it either varies or is constant across a sample set, following the pattern of $T_\mathrm{dep}$ indicated in Fig. \ref{SampleSetTypesGraph}.  In section \ref{DMI} we show that the recorded $P_\mathrm{base}$ determines how strongly the DMI is influenced by $T_\mathrm{dep}$.

\begin{figure}[t]
\includegraphics[scale=0.35]{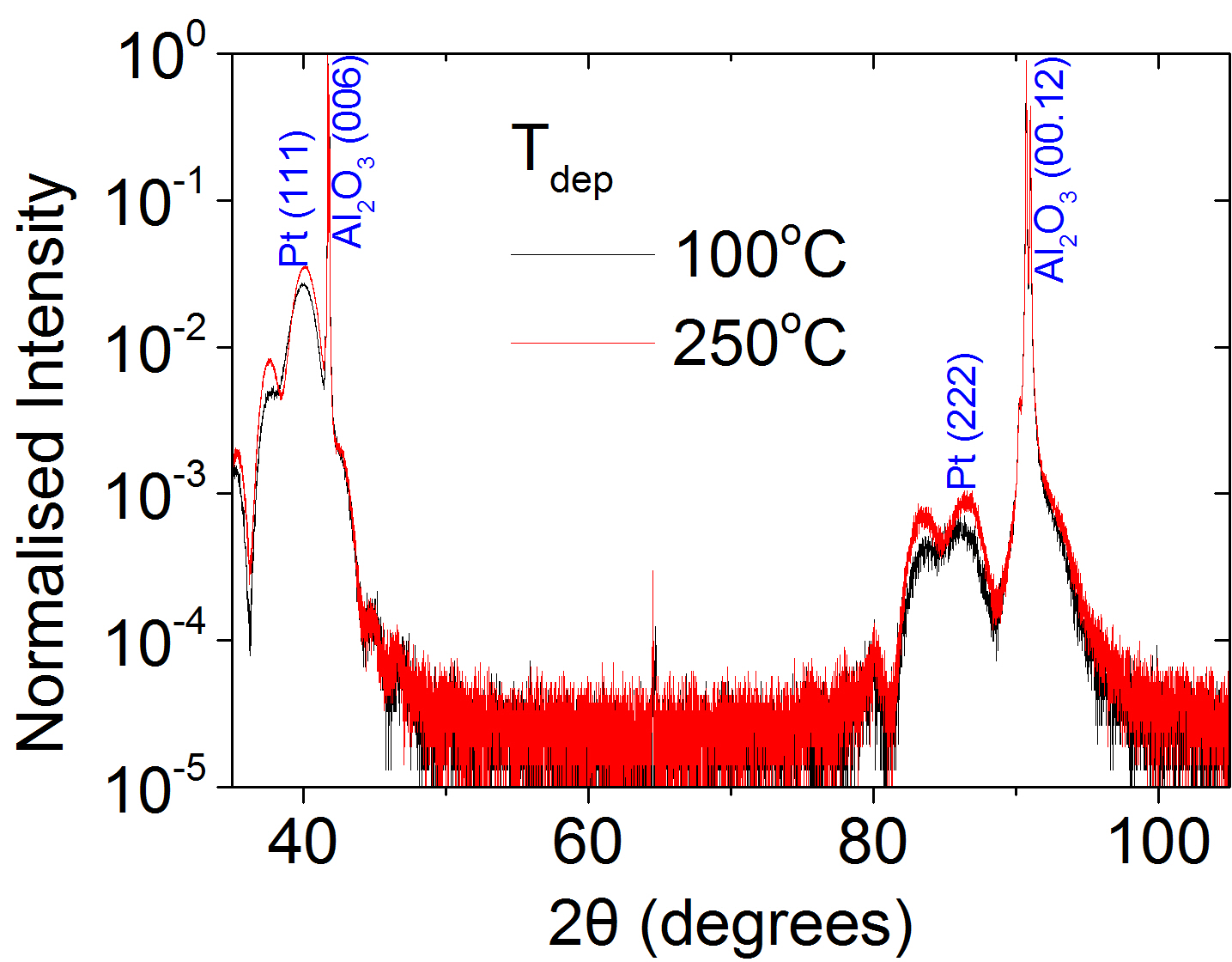}%
\caption{\label{HAXGraph} High angle XRD scans showing the epitaxial nature of the Pt/Co/Pt film for low and high deposition temperatures. The Pt interfaces are of sufficient quality to produce Pendell$\ddot{\mathrm{o}}$sung fringes. The Co peaks are obscured by the Pt peaks. The low intensity peak at 65$^\circ$ is thought to be a higher-order substrate peak.}
\end{figure}

\section{\label{Structural}Structural characterization}

Fig. \ref{HAXGraph} shows high angle X-ray diffraction (XRD) $\mathrm{\theta-2\theta}$ scans for type A samples with the Co and top Pt layer deposited at 100$^\circ$C and 250$^\circ$C, at the low and high end of the $T_\mathrm{dep}$ range that we focus on, measured using $\mathrm{CuK_{\alpha1}}$ X-rays. There are peaks corresponding to fcc Pt (111) that obscure any Co (111) or (0001) peaks; the lack of any further peaks associated with Co or Pt confirms the epitaxial nature of the samples at low and high $T_\mathrm{dep}$. The peaks for the sample deposition at the higher temperature (250$^\circ$C) display better-defined Pendell$\ddot{\mathrm{o}}$sung fringes, indicating a more homogeneous layer thickness between the interfaces of the Pt, consistent with higher atomic mobility during deposition. Since Pt is present at all interfaces in the stack, no information about individual interfaces is gained at this stage.

\begin{figure}[t]
\includegraphics[scale=0.5]{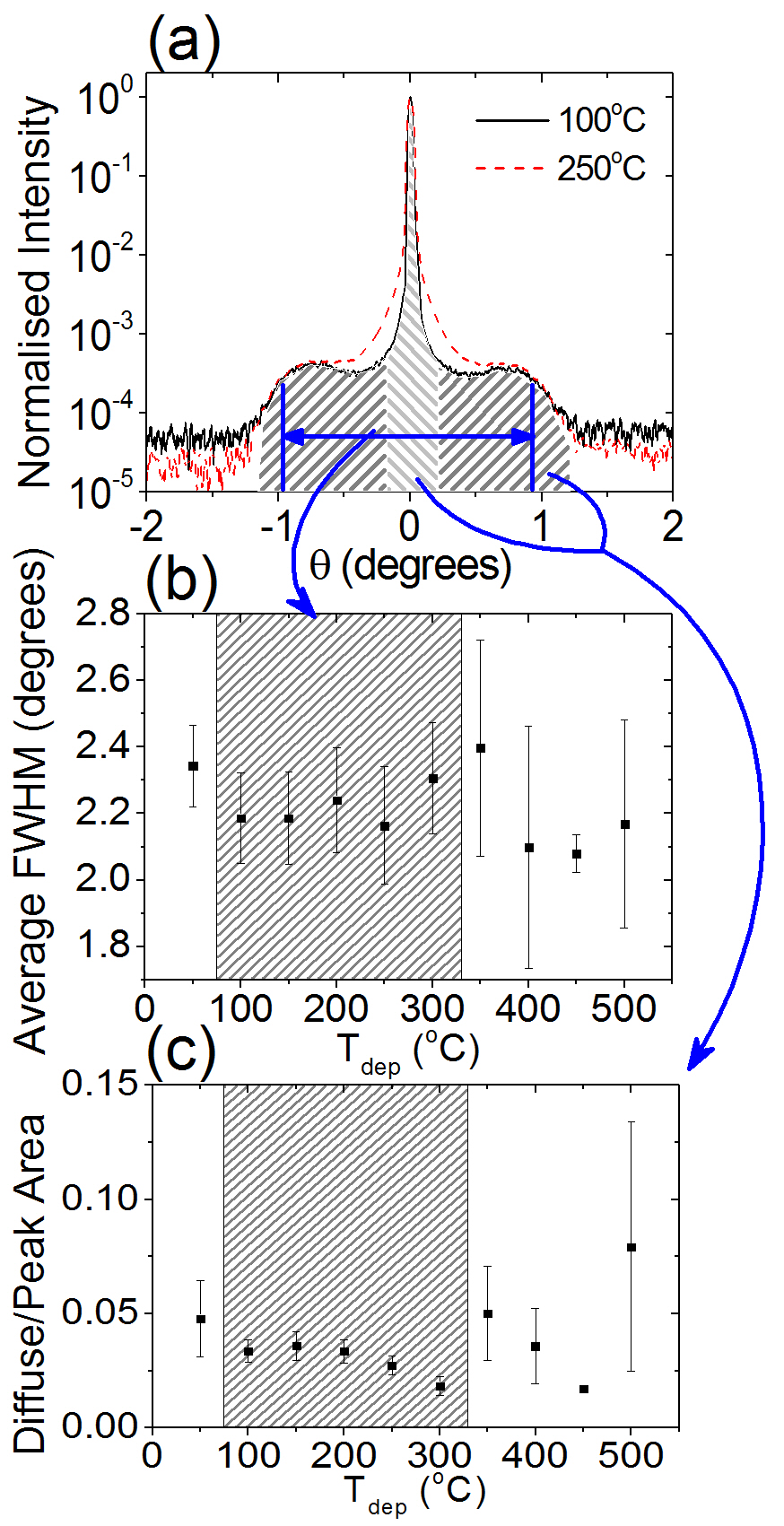}%
\caption{\label{RockingCurvesGraph} Analysis of roughness in Pt/Co/Pt deposited at different temperatures from (a) rocking curves on the first Kiessig fringe. Representative data for low and high $T_\mathrm{dep}$ is shown. The light and dark shaded areas represent the areas under the peak and diffuse wings respectively. (b) FWHM of the diffuse background.(c) Ratio of the area under the diffuse section to that under the peak. The shaded areas in (b) and (c) highlight the main region of interest where the film is epitaxial with distinct layers.}
\end{figure}

To investigate the dependence of the interface roughness on  $T_\mathrm{dep}$, rocking curves (Fig. 3(a))
were measured at the first Kiessig fringe of an X-ray reflectivity scan (e.g. at the 2.7$^\circ$ position of the graph in Fig. 4(a)).
The peak and shoulders of the rocking curve include contributions from all layers, although the surface and bottom-most interfaces are dominant\cite{Sinha88}. This allows the Born approximation to be employed, where multiple reflections are disregarded. Here we define roughness as the collective deviation of atoms from an atomically sharp interface, and intermixing as the deviation of individual atoms. The ratio of diffuse to specular area can be used to assess the  roughness since it is the interfacial roughness which effectively transfers intensity from the specular peak to the diffuse component\cite{Savage91}.

Fig. 3(b) shows no change within error of the FWHM of the diffuse scattering which, when combined with the shape, indicates that the diffuse background is dominated by Yoneda scattering and is therefore defined by the refractive indices of the materials in the multilayer. In Fig. 3(c) the slight decrease of diffuse/specular area in the primary temperature range of 100$^\circ$C to 300$^\circ$C, however, indicates a decrease in the amount of roughness as the deposition temperature increases, probably due to the increasing mobility of the atoms, permitting a lower density of areas of roughness.

\begin{figure}[t]
\includegraphics[scale=0.4]{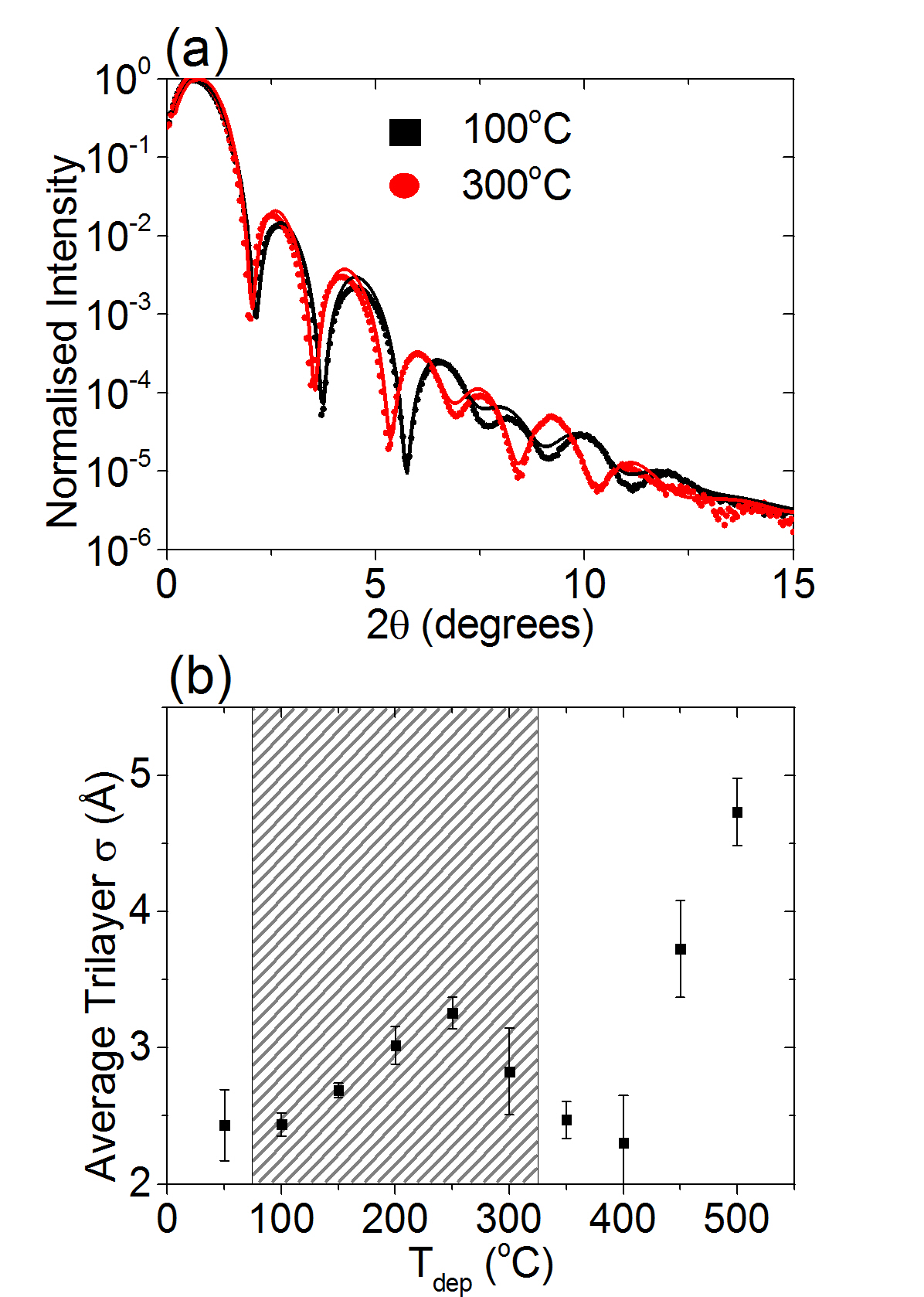}%
\caption{\label{LAXGraph} (a) Low-angle X-ray scans for Pt/Co/Pt deposited at low and high temperatures. The lines are from fits to the data using Bede REFS. (b) $\sigma$ as a function of $T_\mathrm{dep}$, averaged over all samples deposited at each temperature. The shaded area in (b) indicates the main region of interest where the film is epitaxial with distinct layers.}
\end{figure}

X-ray reflectivity scans ($\mathrm{\theta-2\theta}$) at low angles were fitted using Bede REFS\cite{Wormington99} as shown in Fig. \ref{LAXGraph}(a). The sum in quadrature of roughness and intermixing, $\sigma$, was used as an indicator of interface quality (small $\sigma$ indicating a high quality interface), since the wave transfer vector remains normal to the surface in $\mathrm{\theta-2\theta}$ scans making roughness and intermixing indistinguishable. In Fig. 4(b) $\sigma$ averaged over the Pt/Co/Pt structure is plotted as a function of $T_\mathrm{dep}$ for sample sets of type A. The general increase in $\sigma$ with temperature seen in Fig. \ref{LAXGraph}(b) for deposition temperatures 50-250$^\circ$C is opposite to the trend in roughness seen in Fig. \ref{RockingCurvesGraph}(c) so we may deduce that it is predominantly due to an increase in Co-Pt intermixing. For $T_\mathrm{dep}$ from 50-250$^\circ$C, therefore, the Pt/Co/Pt becomes increasingly intermixed until at 250$^\circ$C it reaches the point where the decrease of roughness outweighs any increase in intermixing, as indicated by a decrease in $\sigma$ above 250$^\circ$C.

According to equilibrium phase diagrams\cite{Bolzoni84}, our $T_\mathrm{dep}$ is never quite high enough to produce L1$_\mathrm{0}$ ordered alloys, but above ~400$^\circ$C, they indicate that a disordered fcc CoPt alloy is grown rather than distinct, ordered fcc Co and Pt layers, and we suppose that this is what causes the sharp upturn in $\sigma$ for deposition temperatures  $\geq400^\circ$C.

The structural characterization thus shows that it is possible to have some control over the overall interface quality in epitaxial Pt/Co/Pt via $T_\mathrm{dep}$. While the average $\sigma$ value increases, the roughness decreases slightly as the deposition temperature of the Co and top Pt layer increases from 50 to 300$^\circ$C, from which we deduce that Co-Pt intermixing contributes most to the increase in $\sigma$. We show in sections \ref{Magnetic} and \ref{DMI}, respectively, that this change in interface quality does not significantly affect the perpendicular anisotropy, but that it is linked to changes in the net Dzyaloshinskii-Moriya interaction, indicating that, as $T_\mathrm{dep}$ is increased, the qualities of the top and bottom Co interfaces do not change at the same rate.

\section{\label{Magnetic}Magnetic Characterization}

To investigate the magnetic properties of the type A films, hysteresis loops were measured using the polar magneto-optical Kerr effect (MOKE) with field applied perpendicular to the plane of the film.  As shown in Fig. 5(a), for samples deposited at temperatures in the range 100-250$^\circ$C, the loops are square, indicating that perpendicular anisotropy dominates.  The coercivity, shown in Fig. 5(b), increases as a function of $T_\mathrm{dep}$ in a similar range of deposition temperatures as the interface quality ($1/\sigma$) decreases, as measured in the previous section. We speculate that the intermixing of Co and Pt that increases in the $T_\mathrm{dep}$ range 100-300$^\circ$C leads to an effective broadening and smoothing of the magnetic (rather than chemical) interface that reduces the density and strength of magnetic defects, that in turn leads to larger nucleation fields for reverse domains, and hence larger coercivities. This agrees with our findings of a more homogeneous Co layer thickness and a lower density of areas of roughness at the upper end of the $T_\mathrm{dep}$ range.

\begin{figure}[t]
\includegraphics[scale=0.4]{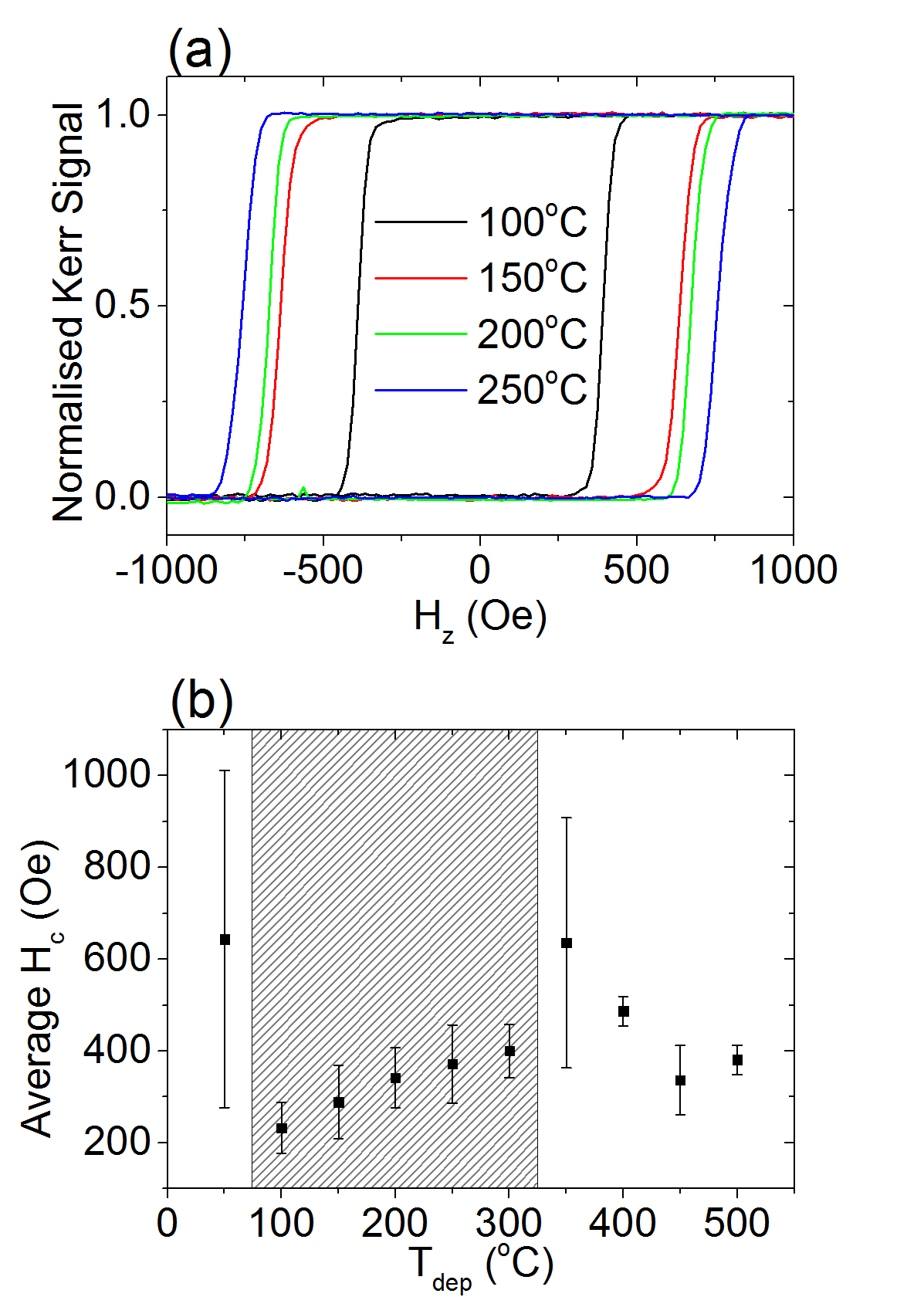}%
\caption{\label{MOKEGraph} (a) Polar MOKE hysteresis loops for Pt/Co/Pt deposited at different temperatures. (b) Coercive field as a function of $T_\mathrm{dep}$, averaged over all samples deposited at each temperature. The shaded area indicates the main region of interest where the film is epitaxial with distinct layers.}
\end{figure}

To determine the perpendicular anisotropy, the sample was rotated through an angle $\theta$ about an in-plane axis in a constant magnetic field while the voltage $V$ associated with the extraordinary Hall effect was measured. The resulting $V-\theta$ data enabled calculation of the effective anisotropy field, $H_\mathrm{K} $\cite{Moon09}. All $H_\mathrm{K}$ values for films of type A deposited at temperatures in the range 100-300$^\circ$C were found to cluster around an average of 14.900 $\pm$ 0.3 kOe. This single value of anisotropy field indicates that any structural change in the sample, e.g. due to interfacial intermixing, is too small to significantly affect the effective anisotropy. 

\begin{figure}[t]
\includegraphics[scale=0.3]{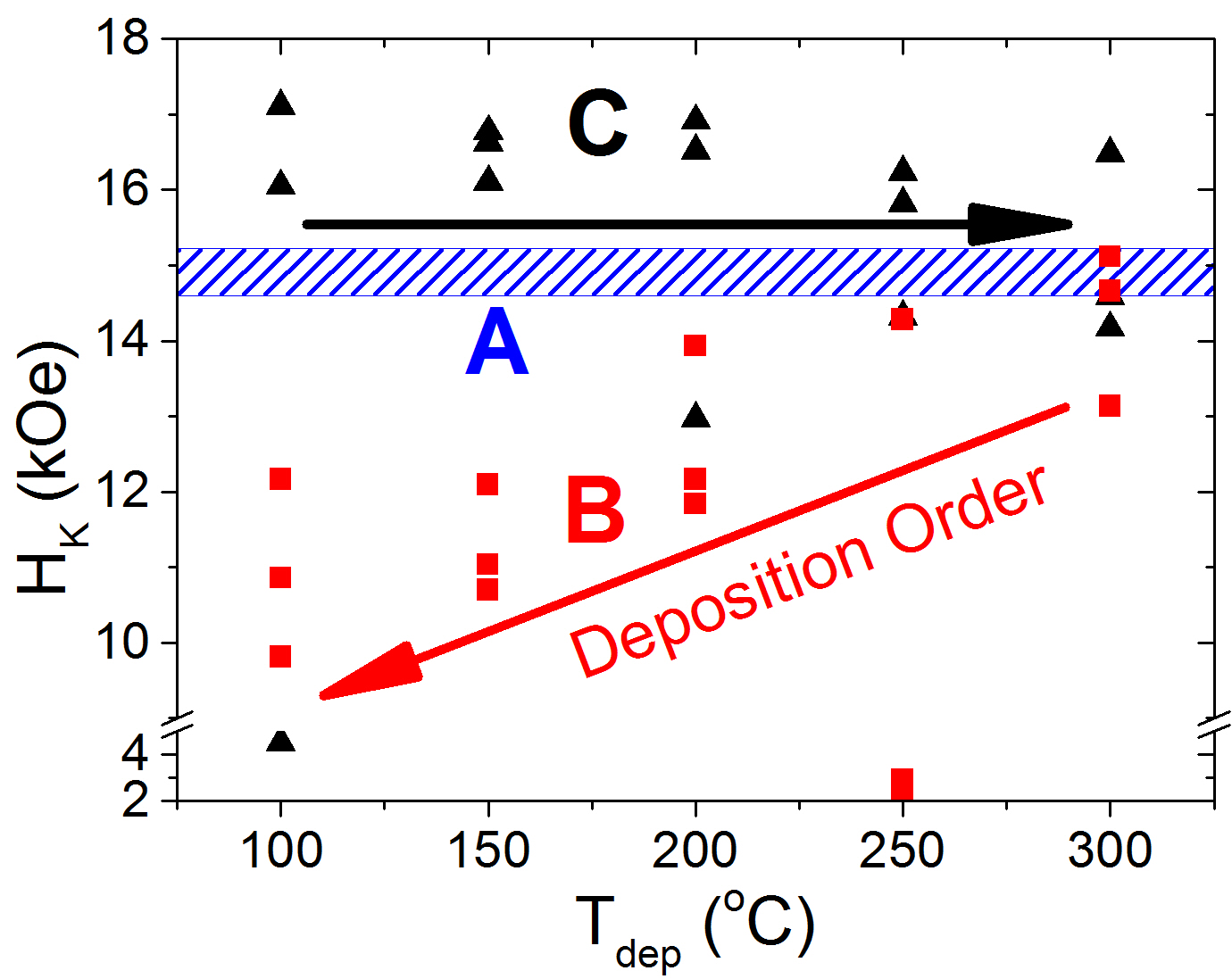}%
\caption{\label{HkGraph} Effective anisotropy field, $H_\mathrm{K}$, for Pt/Co/Pt where the deposition temperature of the top Co/Pt layers was varied (type A), and where only the deposition temperature of the top Pt layer was varied (types B and C). The hatched area indicates $H_\mathrm{K}$ measured for type A films, including uncertainty. The arrows show the order of deposition: 300-100$^\circ$C (type B, squares), and 100-300$^\circ$C (type C, up triangles). Error bars are smaller than or comparable to symbol sizes.}
\end{figure}

To investigate the contributions of the top and bottom Co interfaces to the anisotropy, $H_\mathrm{K}$ was also measured for films of type B and C in which the deposition temperature of only the top Pt layer was varied, and the deposition conditions of the lower interface were kept constant. Fig. \ref{HkGraph} shows that, for films of type B grown in the usual order (hot to cold), the anisotropy field decreases as $T_\mathrm{dep}$ decreases, while for films of type C, grown in reverse (cold to hot), there is very little change in the anisotropy field. Two conclusions may be drawn from this.  The first is that annealing the films at 300$^\circ$C, as occurs for type C, produces a uniform anisotropy field across the set.  (The anisotropy field for type C is 1-1.5 kOe larger than the anisotropy field for type A, suggesting that annealing improves the anisotropy slightly.)  The second is that forming the lower Co interface first at a fixed $T_\mathrm{dep}$, and subsequently forming the upper interface at successively lower temperatures, has the effect of reducing the anisotropy field monotonically from the type C value.  The latter indicates that both upper and lower Co interfaces contribute to the total effective perpendicular anisotropy in all films.  For films of type A, as we have seen, changes in the quality of these interfaces introduced by deposition at different temperatures in the range 100-300$^\circ$C has little effect on the perpendicular anisotropy.  This is useful because the effect of interface quality on DMI can now be investigated with the perpendicular anisotropy effectively kept constant.

\section{\label{DMI}Dzyaloshinskii-Moriya interaction}

\begin{figure*}[!ht]
\includegraphics[scale=0.6]{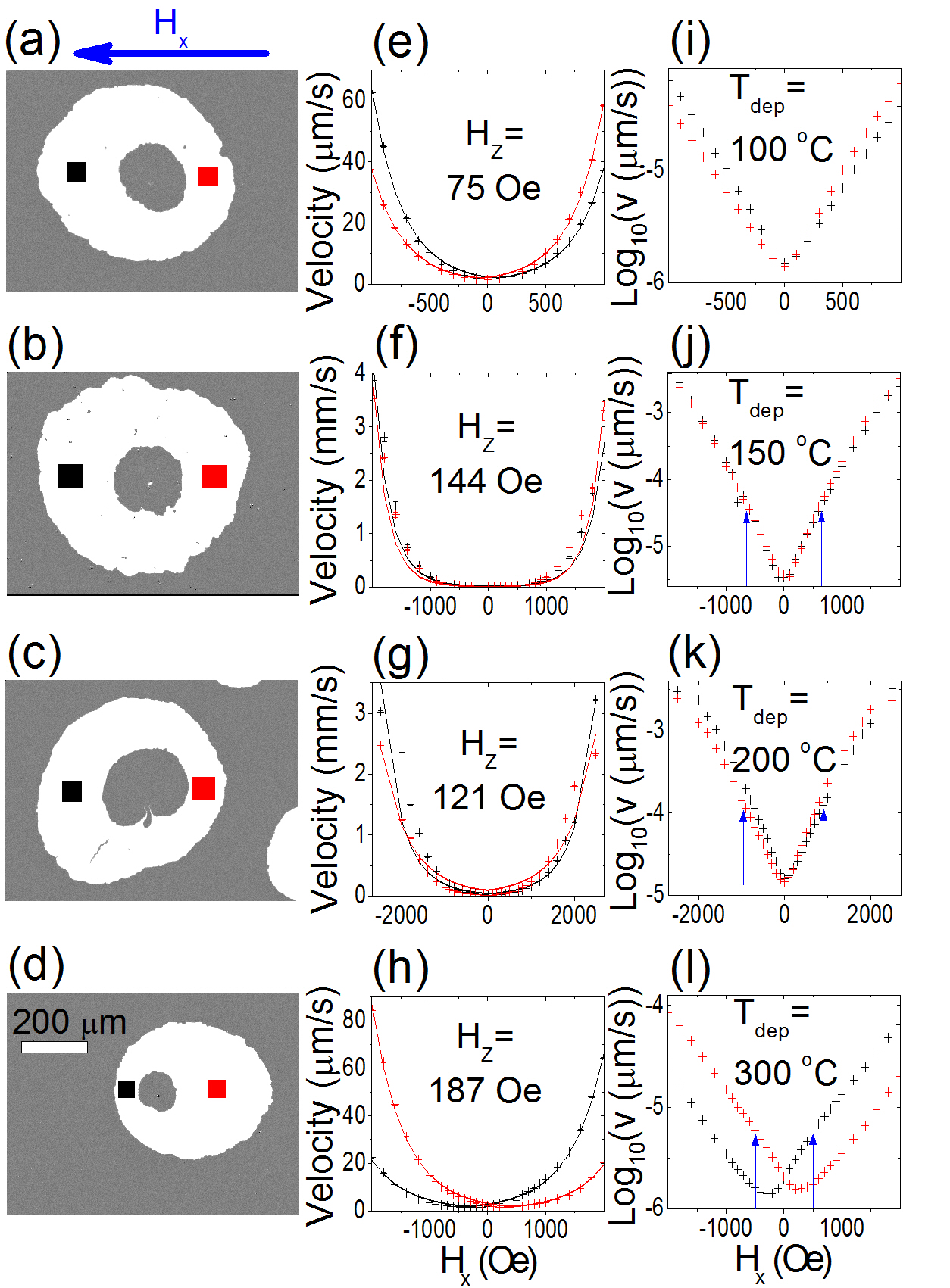}%
\caption{\label{DWVGraph} (a) to (d) Differential mode Kerr microscope images of bubble domains used to determine domain wall displacements in a field of $H_\mathrm{x} =$ -1000 Oe, for a set of Pt/Co/Pt films grown at low base pressure (1.1$\times$10$^{-7}$ Torr) and deposition temperatures indicated for each row. (e) to (h) DW velocities fitted by Eq. 1, and (i) to (l) the logarithm of DW velocity, as a function of $H_\mathrm{x}$. $H_\mathrm{z}$ is the out-of-plane pulsed field used to expand the domain. Black and red correspond to the left- or right-moving wall of the domain, respectively, and the blue arrows indicate the domain wall anisotropy field.}
\end{figure*}

The interfacial Dzyaloshinskii-Moriya interaction (DMI) promotes the formation of chiral N\'{e}el-type domain walls in films with perpendicular anisotropy. The method of expanding a domain in an in-plane field\cite{Hrabec14,Je13} is used to determine the DMI field in Pt/Co/Pt of type A. Kerr imaging is used in the quasi-static regime: a bubble domain is nucleated via an out-of-plane magnetic field pulse, $H_\mathrm{z}$, and a background image is taken. A constant in-plane magnetic field, $H_\mathrm{x}$, is then applied while $H_\mathrm{z}$ is pulsed again. The background image is subtracted from the resulting image to show the domain wall (DW) motion, Fig. \ref{DWVGraph} (a-d), and thus the DW velocity (e-l). The pulse length is of sufficient duration ($>$1 s) that the rise time ($<$33 ms) is negligible. By repeating for various values of $H_\mathrm{x}$ the minimum DW velocity is found, at which point we assume that $H_\mathrm{x} = H_\mathrm{DMI}$, the effective DMI field.  There are reports\cite{Lavrijsen15, Jue15, Lau16} that the DMI might not be the only factor behind the shift of the minimum DW velocity away from $H_\mathrm{x}$=0, but we assume it is the dominant factor here because the DW velocity vs $H_\mathrm{x}$ curves in Fig. 7(e-l) do not exhibit a large asymmetry. The shift in the minimum DW velocity away from $H_\mathrm{x}$=0 is reproduced in a model with $H_\mathrm{DMI}$ as the only term that introduces asymmetry. This model is based on a creep law\cite{Cayssol04, Metaxas07} with the exponential term dependent on the ratio of domain wall energy densities:

\begin{equation} 
\label{creepEq}
v=v_\mathrm{0} \exp[\zeta(\mu_\mathrm{0} H_\mathrm{z} )^{-1/4} ]\mathrm{,}
\end{equation}
where $$\zeta=\zeta_\mathrm{0} [\epsilon(H_\mathrm{x} ) / \epsilon(0) ]^{1/4}\mathrm{.}$$

Here $v_\mathrm{0}$ is the characteristic velocity, $\zeta_\mathrm{0}$ is a scaling constant, and for a domain wall with mixed Bloch-N\'{e}el spin structure the energy density is :

$$\epsilon=\epsilon_\mathrm{0}-\frac{\pi^2 \delta \mu_\mathrm{0}^2  M_\mathrm{s}^2}{8K_\mathrm{D}} (H_\mathrm{x}+H_\mathrm{DMI} )^2\mathrm{.}$$

For a domain wall with pure N\'{e}el spin structure the domain wall energy density is:

$$\epsilon=\epsilon_\mathrm{0}+2K_\mathrm{D} \delta-\pi\delta\mu_\mathrm{0} M_\mathrm{s} |H_\mathrm{x}+H_\mathrm{DMI} |\mathrm{,}$$
where the Bloch wall energy density $\epsilon_\mathrm{0}=2\pi\sqrt{A K_\mathrm{0} }$, the exchange stiffness $A$ = 16 $\times$ 10$^{-12}$ J/m and $K_\mathrm{0}$ is the effective anisotropy, used as a fitting parameter; $\delta=\sqrt{A / K_\mathrm{0} }$ is the domain wall width; $M_\mathrm{s}$ = 1.1 $\times$ 10$^6$A/m is the saturation magnetization; and $K_\mathrm{D}=(N_\mathrm{x} \mu_\mathrm{0} M_\mathrm{s}^2) / 2$ is the domain wall anisotropy where the demagnetizing factor $N_\mathrm{x} = \ln(⁡2)t / (\pi\delta)$ with $t$ the thickness of the ferromagnetic film. Values of saturation magnetization and exchange stiffness were taken from previous work\cite{Hrabec14, Thiaville12}.

Measuring DW velocity as a function of $H_\mathrm{z}$ with no applied $H_\mathrm{x}$ we found a linear relationship between $\ln v$ and $H_\mathrm{z}^{-1/4}$, confirming that the DW motion in the Pt/Co/Pt films was, indeed, in the creep regime. Fitting to $\ln v$ vs. $H_\mathrm{z}^{-1/4}$, $v_\mathrm{0}$ and $\zeta_\mathrm{0}$ were determined, leaving just $H_\mathrm{DMI}$ and $K_\mathrm{0}$ as fitting parameters for velocities measured as a function of $H_\mathrm{x}$, shown in Fig. \ref{DWVGraph} (e-h). Values of $K_\mathrm{0}$ determined from fits to the data in Fig \ref{DWVGraph} (e-h) ranged from approximately 30-60 $\mathrm{kJ/m}^{3}$, where the upper bound is the effective anisotropy found from the average $H_\mathrm{K}$ measured in section \ref{Magnetic}. The spread in $K_\mathrm{0}$ may arise as a result of the range of local anisotropies probed by the moving domain wall, as opposed to the single value of anisotropy field measured for the samples as a whole. $H_\mathrm{DMI}$, however, is more reliably determined as it controls the asymmetry of the fit.

Data shown in Fig. \ref{DWVGraph} for a set of films grown at a base pressure of 1.1$\times$10$^{-7}$ Torr display the asymmetric expansion of the bubble domains, (a-d), as well as the shift in minimum of the DW velocity, (e-h), (i-l). A slight change of gradient is apparent when log(v) is plotted as a function of $H_\mathrm{x}$, indicated by the arrows in (i-l), and in other literature has been attributed to chiral damping\cite{Jue15}. However, the field at which this change of gradient occurs is at roughly the same distance from the minimum as the magnitude of the domain wall shape anisotropy field (310 $\pm$ 90 Oe), indicating that at this point the DW spin structure changes from mixed Bloch-N\'{e}el to pure N\'{e}el. Furthermore, the net DMI fields were found to range between -90 and 276 Oe, and so $H_\mathrm{DMI}$ alone is not sufficient to bring the domain walls into a pure N\'{e}el configuration in any of the films, meaning that at $H_\mathrm{x}=0$ the domain wall spin structure is mixed Bloch-N\'{e}el.  The DW velocities in Fig. 7(e-h) were fitted by Eq. 1 with the DW energy density (mixed Bloch-N\'{e}el, or pure N\'{e}el) chosen according to whether $H_\mathrm{x}$ was greater than or less than the DW shape anisotropy field.

\begin{figure}[t]
\includegraphics[scale=0.3]{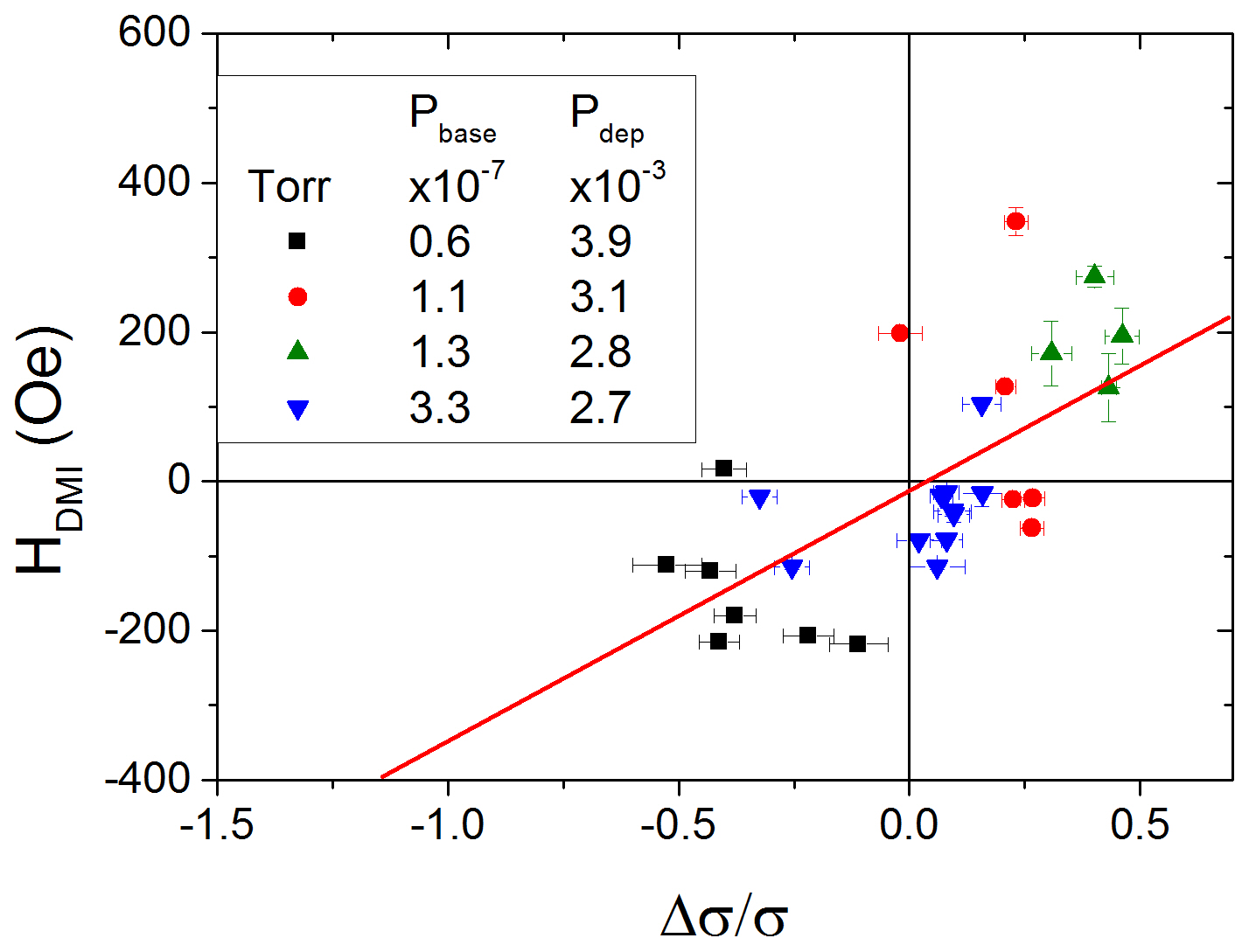}%
\caption{\label{HDMISGraph} Net $H_\mathrm{DMI}$ as a function of $\Delta\sigma/\sigma$, the difference between the top Co interface $\sigma$ and the bottom Co interface $\sigma$ for epitaxial Pt/Co/Pt deposited at temperatures in the range 100-300$^\circ$C, normalized by the total $\sigma$. A positive $\sigma$ difference corresponds to the lower Co interface being a higher quality than the upper. The errors associated with $\Delta\sigma/\sigma$ relate to the suitability of the $\sigma$ values to the fitted X-ray reflectivity models so act as a lower bound of uncertainty due to the deviations of the fits. The measured values of $K_\mathrm{0}$ were used in the conversion of $H_\mathrm{DMI}$ to D.}
\end{figure}

Fig. \ref{HDMISGraph} shows $H_\mathrm{DMI}$ for sets of films grown at a range of base pressures from 0.6-3.3$\times$10$^{-7}$ Torr plotted against the normalised difference in $\sigma$ for the upper and lower Co interfaces.  The difference in $\sigma$ is a measure of the difference in quality of the upper and lower Co interfaces. Measuring the net DMI field for several sets of samples permits a correlation to emerge, as the  DMI is exquisitely sensitive to the difference in quality of the upper and lower Co interfaces. The difference in $\sigma$ is normalized by the total $\sigma$ to highlight any dependence on the interface quality difference, independent of any changes in total $\sigma$.


Fig. \ref{HDMISGraph} shows that the net DMI in epitaxial Pt/Co/Pt increases from zero to positive values when the quality of the top interface decreases relative to the bottom, and decreases from zero to negative values when the quality of bottom interface decreases relative to the top. A linear fit to the data yields a Pearson's $r$ correlation coefficient of 0.65. The spread of data points is due to the  uncertainty in $\sigma$ for individual interfaces determined from fitting low-angle X-ray scans of such thin films. 

The conclusion that may be drawn from Fig. \ref{HDMISGraph} is that the net DMI field in epitaxial Pt/Co/Pt can change by up to 340 $\pm$ 60 Oe depending on the deposition conditions. For less ordered, polycrystalline Pt/Co/Pt, therefore, it is no surprise that larger DMI fields may be obtained, e.g. Franken et al\cite{Franken14} obtained $H_\mathrm{DMI}$=370 $\pm$ 10 Oe in a polycrystalline stack with a thinner Co layer (4 \r{A}) and Hrabec et al.\cite{Hrabec14} obtained $H_\mathrm{DMI}$\mytilde-1000 Oe for films with a similar Co thickness but polycrystalline and with a different Pt thickness. Na$\ddot{\i}$vely mapping $H_\mathrm{DMI}$\mytilde-1000 Oe onto Fig. \ref{HDMISGraph} yields a difference in $\Delta\sigma/\sigma$ of -2.9, which would mean that the difference in interface quality is greater than the total interface quality, and suggests that there is another factor at work here, possibly related to the polycrystallinity.


\begin{figure}[t]
\includegraphics[scale=0.3]{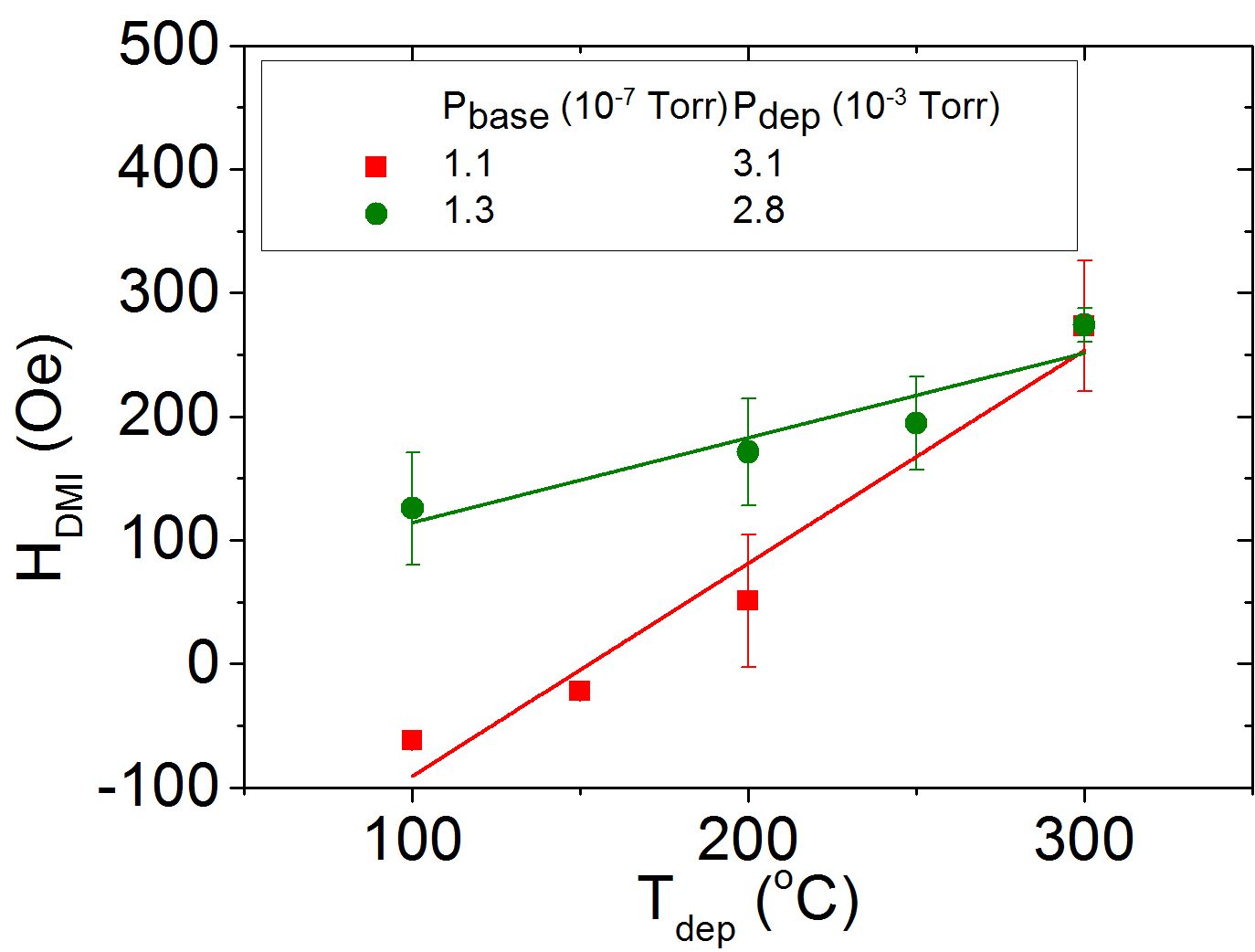}%
\caption{\label{HDMITGraph} Net $H_\mathrm{DMI}$ as a function of $T_\mathrm{dep}$ for epitaxial Pt/Co/Pt where growth was initiated at low base pressures.}
\end{figure}

For samples of type A where growth was initiated at sufficiently low base pressure, $H_\mathrm{DMI}$ increases monotonically as a function of $T_\mathrm{dep}$, as shown in Fig. \ref{HDMITGraph}.  Using the prior finding that the lower Co interface contributes a positive $H_\mathrm{DMI}$\cite{Hrabec14}, and the conclusion drawn earlier that a higher quality (smaller $\sigma$) interface contributes a larger $H_\mathrm{DMI}$, the increase of $H_\mathrm{DMI}$ with $T_\mathrm{dep}$ may be interpreted as follows. As $T_\mathrm{dep}$ increases, the quality of both interfaces, but particularly the upper, decreases until annealing dominates and their quality improves; the lower at a greater rate than the upper. This difference in interface quality introduces structural inversion asymmetry, necessary for a net DMI field to occur, increasing as the contribution from the top interface is reduced, as seen in Fig. \ref{HDMITGraph}. As the Co interfaces become of a similar quality, the DMI contributions from the top and bottom cancel, causing no net DMI. If the bottom interface is of a lower quality than the top, the dominant contribution switches and the effective DMI field becomes negative. This shows that if the base pressure is sufficiently low, substrate temperature may be used to linearly adjust the DMI.


\section{\label{Conclusion}Conclusion}

In conclusion, changing the temperature during deposition of the top Co/Pt layers in an epitaxially sputtered Pt/Co/Pt system significantly affects the Co interface quality. The difference between the quality of the top interface and the lower interface introduces structural inversion asymmetry which results in a net DMI field. This difference is altered with deposition temperature as the interfaces change quality at different rates. This shows that, for sufficiently good base pressures, the substrate temperature may be used to fine-tune the DMI in epitaxial samples.

This work was supported by the U.K. EPSRC (Grants No. EP/I011668/1, No. EP/K003127/1 and No. EP/M000923/1).

\bibliography{DMI tuning paper- LaTeX- CHM-6_TAM4_AWJW}

\begin{thebibliography}{37}%
\makeatletter
\providecommand \@ifxundefined [1]{%
 \@ifx{#1\undefined}
}%
\providecommand \@ifnum [1]{%
 \ifnum #1\expandafter \@firstoftwo
 \else \expandafter \@secondoftwo
 \fi
}%
\providecommand \@ifx [1]{%
 \ifx #1\expandafter \@firstoftwo
 \else \expandafter \@secondoftwo
 \fi
}%
\providecommand \natexlab [1]{#1}%
\providecommand \enquote  [1]{``#1''}%
\providecommand \bibnamefont  [1]{#1}%
\providecommand \bibfnamefont [1]{#1}%
\providecommand \citenamefont [1]{#1}%
\providecommand \href@noop [0]{\@secondoftwo}%
\providecommand \href [0]{\begingroup \@sanitize@url \@href}%
\providecommand \@href[1]{\@@startlink{#1}\@@href}%
\providecommand \@@href[1]{\endgroup#1\@@endlink}%
\providecommand \@sanitize@url [0]{\catcode `\\12\catcode `\$12\catcode
  `\&12\catcode `\#12\catcode `\^12\catcode `\_12\catcode `\%12\relax}%
\providecommand \@@startlink[1]{}%
\providecommand \@@endlink[0]{}%
\providecommand \url  [0]{\begingroup\@sanitize@url \@url }%
\providecommand \@url [1]{\endgroup\@href {#1}{\urlprefix }}%
\providecommand \urlprefix  [0]{URL }%
\providecommand \Eprint [0]{\href }%
\providecommand \doibase [0]{http://dx.doi.org/}%
\providecommand \selectlanguage [0]{\@gobble}%
\providecommand \bibinfo  [0]{\@secondoftwo}%
\providecommand \bibfield  [0]{\@secondoftwo}%
\providecommand \translation [1]{[#1]}%
\providecommand \BibitemOpen [0]{}%
\providecommand \bibitemStop [0]{}%
\providecommand \bibitemNoStop [0]{.\EOS\space}%
\providecommand \EOS [0]{\spacefactor3000\relax}%
\providecommand \BibitemShut  [1]{\csname bibitem#1\endcsname}%
\let\auto@bib@innerbib\@empty
\bibitem [{\citenamefont {Pizzini}\ \emph {et~al.}(2014)\citenamefont
  {Pizzini}, \citenamefont {Vogel}, \citenamefont {Rohart}, \citenamefont
  {Buda-Prejbeanu}, \citenamefont {Ju\'{e}}, \citenamefont {Boulle},
  \citenamefont {Miron}, \citenamefont {Safeer}, \citenamefont {Auffret},
  \citenamefont {Gaudin},\ and\ \citenamefont {Thiaville}}]{Pizzini14}%
  \BibitemOpen
  \bibfield  {author} {\bibinfo {author} {\bibfnamefont {S.}~\bibnamefont
  {Pizzini}}, \bibinfo {author} {\bibfnamefont {J.}~\bibnamefont {Vogel}},
  \bibinfo {author} {\bibfnamefont {S.}~\bibnamefont {Rohart}}, \bibinfo
  {author} {\bibfnamefont {L.~D.}\ \bibnamefont {Buda-Prejbeanu}}, \bibinfo
  {author} {\bibfnamefont {E.}~\bibnamefont {Ju\'{e}}}, \bibinfo {author}
  {\bibfnamefont {O.}~\bibnamefont {Boulle}}, \bibinfo {author} {\bibfnamefont
  {I.~M.}\ \bibnamefont {Miron}}, \bibinfo {author} {\bibfnamefont {C.~K.}\
  \bibnamefont {Safeer}}, \bibinfo {author} {\bibfnamefont {S.}~\bibnamefont
  {Auffret}}, \bibinfo {author} {\bibfnamefont {G.}~\bibnamefont {Gaudin}}, \
  and\ \bibinfo {author} {\bibfnamefont {A.}~\bibnamefont {Thiaville}},\
  }\href@noop {} {\bibfield  {journal} {\bibinfo  {journal} {Phys. Rev. Lett.}\
  }\textbf {\bibinfo {volume} {113}},\ \bibinfo {pages} {047203} (\bibinfo
  {year} {2014})}\BibitemShut {NoStop}%
\bibitem [{\citenamefont {Belmeguenai}\ \emph {et~al.}(2015)\citenamefont
  {Belmeguenai}, \citenamefont {Adam}, \citenamefont {Roussign\'{e}},
  \citenamefont {Eimer}, \citenamefont {Devolder}, \citenamefont {Kim},
  \citenamefont {Cherif}, \citenamefont {Stashkevich},\ and\ \citenamefont
  {Thiaville}}]{Belmeguenai15}%
  \BibitemOpen
  \bibfield  {author} {\bibinfo {author} {\bibfnamefont {M.}~\bibnamefont
  {Belmeguenai}}, \bibinfo {author} {\bibfnamefont {J.-P.}\ \bibnamefont
  {Adam}}, \bibinfo {author} {\bibfnamefont {Y.}~\bibnamefont {Roussign\'{e}}},
  \bibinfo {author} {\bibfnamefont {S.}~\bibnamefont {Eimer}}, \bibinfo
  {author} {\bibfnamefont {T.}~\bibnamefont {Devolder}}, \bibinfo {author}
  {\bibfnamefont {J.-V.}\ \bibnamefont {Kim}}, \bibinfo {author} {\bibfnamefont
  {S.~M.}\ \bibnamefont {Cherif}}, \bibinfo {author} {\bibfnamefont
  {A.}~\bibnamefont {Stashkevich}}, \ and\ \bibinfo {author} {\bibfnamefont
  {A.}~\bibnamefont {Thiaville}},\ }\href@noop {} {\bibfield  {journal}
  {\bibinfo  {journal} {Phys. Rev. B}\ }\textbf {\bibinfo {volume} {91}},\
  \bibinfo {pages} {180405(R)} (\bibinfo {year} {2015})}\BibitemShut {NoStop}%
\bibitem [{\citenamefont {Benitez}\ \emph {et~al.}(2015)\citenamefont
  {Benitez}, \citenamefont {Hrabec}, \citenamefont {Mihai}, \citenamefont
  {Moore}, \citenamefont {Burnell}, \citenamefont {McGrouther}, \citenamefont
  {Marrows},\ and\ \citenamefont {McVitie}}]{Benitez15}%
  \BibitemOpen
  \bibfield  {author} {\bibinfo {author} {\bibfnamefont {M.~J.}\ \bibnamefont
  {Benitez}}, \bibinfo {author} {\bibfnamefont {A.}~\bibnamefont {Hrabec}},
  \bibinfo {author} {\bibfnamefont {A.~P.}\ \bibnamefont {Mihai}}, \bibinfo
  {author} {\bibfnamefont {T.~A.}\ \bibnamefont {Moore}}, \bibinfo {author}
  {\bibfnamefont {G.}~\bibnamefont {Burnell}}, \bibinfo {author} {\bibfnamefont
  {D.}~\bibnamefont {McGrouther}}, \bibinfo {author} {\bibfnamefont {C.~H.}\
  \bibnamefont {Marrows}}, \ and\ \bibinfo {author} {\bibfnamefont
  {S.}~\bibnamefont {McVitie}},\ }\href@noop {} {\bibfield  {journal} {\bibinfo
   {journal} {Nat. Commun.}\ }\textbf {\bibinfo {volume} {6}},\ \bibinfo
  {pages} {8957} (\bibinfo {year} {2015})}\BibitemShut {NoStop}%
\bibitem [{\citenamefont {Hrabec}\ \emph {et~al.}(2014)\citenamefont {Hrabec},
  \citenamefont {Porter}, \citenamefont {Wells}, \citenamefont {Benitez},
  \citenamefont {Burnell}, \citenamefont {McVitie}, \citenamefont {McGrouther},
  \citenamefont {Moore},\ and\ \citenamefont {Marrows}}]{Hrabec14}%
  \BibitemOpen
  \bibfield  {author} {\bibinfo {author} {\bibfnamefont {A.}~\bibnamefont
  {Hrabec}}, \bibinfo {author} {\bibfnamefont {N.~A.}\ \bibnamefont {Porter}},
  \bibinfo {author} {\bibfnamefont {A.}~\bibnamefont {Wells}}, \bibinfo
  {author} {\bibfnamefont {M.~J.}\ \bibnamefont {Benitez}}, \bibinfo {author}
  {\bibfnamefont {G.}~\bibnamefont {Burnell}}, \bibinfo {author} {\bibfnamefont
  {S.}~\bibnamefont {McVitie}}, \bibinfo {author} {\bibfnamefont
  {D.}~\bibnamefont {McGrouther}}, \bibinfo {author} {\bibfnamefont {T.~A.}\
  \bibnamefont {Moore}}, \ and\ \bibinfo {author} {\bibfnamefont {C.~H.}\
  \bibnamefont {Marrows}},\ }\href@noop {} {\bibfield  {journal} {\bibinfo
  {journal} {Phys. Rev. B}\ }\textbf {\bibinfo {volume} {90}},\ \bibinfo
  {pages} {020402(R)} (\bibinfo {year} {2014})}\BibitemShut {NoStop}%
\bibitem [{\citenamefont {Yang}\ \emph {et~al.}(2016)\citenamefont {Yang},
  \citenamefont {Boulle}, \citenamefont {Cros}, \citenamefont {Fert},\ and\
  \citenamefont {Chshiev}}]{Yang16}%
  \BibitemOpen
  \bibfield  {author} {\bibinfo {author} {\bibfnamefont {H.}~\bibnamefont
  {Yang}}, \bibinfo {author} {\bibfnamefont {O.}~\bibnamefont {Boulle}},
  \bibinfo {author} {\bibfnamefont {V.}~\bibnamefont {Cros}}, \bibinfo {author}
  {\bibfnamefont {A.}~\bibnamefont {Fert}}, \ and\ \bibinfo {author}
  {\bibfnamefont {M.}~\bibnamefont {Chshiev}},\ }\href@noop {} {\bibfield
  {journal} {\bibinfo  {journal} {arXiv}\ }\textbf {\bibinfo {volume} {1603}},\
  \bibinfo {pages} {01847v1} (\bibinfo {year} {2016})}\BibitemShut {NoStop}%
\bibitem [{\citenamefont {Koyama}\ \emph {et~al.}(2011)\citenamefont {Koyama},
  \citenamefont {Chiba}, \citenamefont {Ueda}, \citenamefont {Kondou},
  \citenamefont {Tanigawa}, \citenamefont {Fukami}, \citenamefont {Suzuki},
  \citenamefont {Ohshima}, \citenamefont {Ishiwata}, \citenamefont {Nakatani},
  \citenamefont {Kobayashi},\ and\ \citenamefont {Ono}}]{Koyama11}%
  \BibitemOpen
  \bibfield  {author} {\bibinfo {author} {\bibfnamefont {T.}~\bibnamefont
  {Koyama}}, \bibinfo {author} {\bibfnamefont {D.}~\bibnamefont {Chiba}},
  \bibinfo {author} {\bibfnamefont {K.}~\bibnamefont {Ueda}}, \bibinfo {author}
  {\bibfnamefont {K.}~\bibnamefont {Kondou}}, \bibinfo {author} {\bibfnamefont
  {H.}~\bibnamefont {Tanigawa}}, \bibinfo {author} {\bibfnamefont
  {S.}~\bibnamefont {Fukami}}, \bibinfo {author} {\bibfnamefont
  {T.}~\bibnamefont {Suzuki}}, \bibinfo {author} {\bibfnamefont
  {N.}~\bibnamefont {Ohshima}}, \bibinfo {author} {\bibfnamefont
  {N.}~\bibnamefont {Ishiwata}}, \bibinfo {author} {\bibfnamefont
  {Y.}~\bibnamefont {Nakatani}}, \bibinfo {author} {\bibfnamefont
  {K.}~\bibnamefont {Kobayashi}}, \ and\ \bibinfo {author} {\bibfnamefont
  {T.}~\bibnamefont {Ono}},\ }\href@noop {} {\bibfield  {journal} {\bibinfo
  {journal} {Nat. Mater.}\ }\textbf {\bibinfo {volume} {10}},\ \bibinfo {pages}
  {194} (\bibinfo {year} {2011})}\BibitemShut {NoStop}%
\bibitem [{\citenamefont {Ryu}\ \emph {et~al.}(2013)\citenamefont {Ryu},
  \citenamefont {Thomas}, \citenamefont {Yang},\ and\ \citenamefont
  {Parkin}}]{Ryu13}%
  \BibitemOpen
  \bibfield  {author} {\bibinfo {author} {\bibfnamefont {K.-S.}\ \bibnamefont
  {Ryu}}, \bibinfo {author} {\bibfnamefont {L.}~\bibnamefont {Thomas}},
  \bibinfo {author} {\bibfnamefont {S.-H.}\ \bibnamefont {Yang}}, \ and\
  \bibinfo {author} {\bibfnamefont {S.}~\bibnamefont {Parkin}},\ }\href@noop {}
  {\bibfield  {journal} {\bibinfo  {journal} {Nat. Nanotechnol.}\ }\textbf
  {\bibinfo {volume} {8}},\ \bibinfo {pages} {527} (\bibinfo {year}
  {2013})}\BibitemShut {NoStop}%
\bibitem [{\citenamefont {Chen}\ \emph {et~al.}(2013)\citenamefont {Chen},
  \citenamefont {Ma}, \citenamefont {N'Diaye}, \citenamefont {Kwon},
  \citenamefont {Won}, \citenamefont {Wu},\ and\ \citenamefont
  {Schmid}}]{Chen13}%
  \BibitemOpen
  \bibfield  {author} {\bibinfo {author} {\bibfnamefont {G.}~\bibnamefont
  {Chen}}, \bibinfo {author} {\bibfnamefont {T.}~\bibnamefont {Ma}}, \bibinfo
  {author} {\bibfnamefont {A.~T.}\ \bibnamefont {N'Diaye}}, \bibinfo {author}
  {\bibfnamefont {H.}~\bibnamefont {Kwon}}, \bibinfo {author} {\bibfnamefont
  {C.}~\bibnamefont {Won}}, \bibinfo {author} {\bibfnamefont {Y.}~\bibnamefont
  {Wu}}, \ and\ \bibinfo {author} {\bibfnamefont {A.~K.}\ \bibnamefont
  {Schmid}},\ }\href@noop {} {\bibfield  {journal} {\bibinfo  {journal} {Nat.
  Commun.}\ }\textbf {\bibinfo {volume} {4}},\ \bibinfo {pages} {2671}
  (\bibinfo {year} {2013})}\BibitemShut {NoStop}%
\bibitem [{\citenamefont {Je}\ \emph {et~al.}(2013)\citenamefont {Je},
  \citenamefont {Kim}, \citenamefont {Yoo}, \citenamefont {Min}, \citenamefont
  {Lee},\ and\ \citenamefont {Choe}}]{Je13}%
  \BibitemOpen
  \bibfield  {author} {\bibinfo {author} {\bibfnamefont {S.-G.}\ \bibnamefont
  {Je}}, \bibinfo {author} {\bibfnamefont {D.-H.}\ \bibnamefont {Kim}},
  \bibinfo {author} {\bibfnamefont {S.-C.}\ \bibnamefont {Yoo}}, \bibinfo
  {author} {\bibfnamefont {B.-C.}\ \bibnamefont {Min}}, \bibinfo {author}
  {\bibfnamefont {K.-J.}\ \bibnamefont {Lee}}, \ and\ \bibinfo {author}
  {\bibfnamefont {S.-B.}\ \bibnamefont {Choe}},\ }\href@noop {} {\bibfield
  {journal} {\bibinfo  {journal} {Phys. Rev. B}\ }\textbf {\bibinfo {volume}
  {88}},\ \bibinfo {pages} {214401} (\bibinfo {year} {2013})}\BibitemShut
  {NoStop}%
\bibitem [{\citenamefont {Lavrijsen}\ \emph {et~al.}(2015)\citenamefont
  {Lavrijsen}, \citenamefont {Hartmann}, \citenamefont {van~den Brink},
  \citenamefont {Yin}, \citenamefont {Barcones}, \citenamefont {Duine},
  \citenamefont {Verheijen}, \citenamefont {Swagten},\ and\ \citenamefont
  {Koopmans}}]{Lavrijsen15}%
  \BibitemOpen
  \bibfield  {author} {\bibinfo {author} {\bibfnamefont {R.}~\bibnamefont
  {Lavrijsen}}, \bibinfo {author} {\bibfnamefont {D.~M.~F.}\ \bibnamefont
  {Hartmann}}, \bibinfo {author} {\bibfnamefont {A.}~\bibnamefont {van~den
  Brink}}, \bibinfo {author} {\bibfnamefont {Y.}~\bibnamefont {Yin}}, \bibinfo
  {author} {\bibfnamefont {B.}~\bibnamefont {Barcones}}, \bibinfo {author}
  {\bibfnamefont {R.~A.}\ \bibnamefont {Duine}}, \bibinfo {author}
  {\bibfnamefont {M.~A.}\ \bibnamefont {Verheijen}}, \bibinfo {author}
  {\bibfnamefont {H.~J.~M.}\ \bibnamefont {Swagten}}, \ and\ \bibinfo {author}
  {\bibfnamefont {B.}~\bibnamefont {Koopmans}},\ }\href@noop {} {\bibfield
  {journal} {\bibinfo  {journal} {Phys. Rev. B}\ }\textbf {\bibinfo {volume}
  {91}},\ \bibinfo {pages} {104414} (\bibinfo {year} {2015})}\BibitemShut
  {NoStop}%
\bibitem [{\citenamefont {Va\v{n}atka}\ \emph {et~al.}(2015)\citenamefont
  {Va\v{n}atka}, \citenamefont {Rojas-S\'{a}nchez}, \citenamefont {Vogel},
  \citenamefont {Bonfim}, \citenamefont {Belmeguenai}, \citenamefont
  {Roussign\'{e}}, \citenamefont {Stashkevich}, \citenamefont {Thiaville},\
  and\ \citenamefont {Pizzini}}]{Vanatka15}%
  \BibitemOpen
  \bibfield  {author} {\bibinfo {author} {\bibfnamefont {M.}~\bibnamefont
  {Va\v{n}atka}}, \bibinfo {author} {\bibfnamefont {J.-C.}\ \bibnamefont
  {Rojas-S\'{a}nchez}}, \bibinfo {author} {\bibfnamefont {J.}~\bibnamefont
  {Vogel}}, \bibinfo {author} {\bibfnamefont {M.}~\bibnamefont {Bonfim}},
  \bibinfo {author} {\bibfnamefont {M.}~\bibnamefont {Belmeguenai}}, \bibinfo
  {author} {\bibfnamefont {Y.}~\bibnamefont {Roussign\'{e}}}, \bibinfo {author}
  {\bibfnamefont {A.}~\bibnamefont {Stashkevich}}, \bibinfo {author}
  {\bibfnamefont {A.}~\bibnamefont {Thiaville}}, \ and\ \bibinfo {author}
  {\bibfnamefont {S.}~\bibnamefont {Pizzini}},\ }\href@noop {} {\bibfield
  {journal} {\bibinfo  {journal} {J. Phys.: Condens. Matter}\ }\textbf
  {\bibinfo {volume} {27}},\ \bibinfo {pages} {326002} (\bibinfo {year}
  {2015})}\BibitemShut {NoStop}%
\bibitem [{\citenamefont {Ummelen}\ \emph {et~al.}(2015)\citenamefont
  {Ummelen}, \citenamefont {Han}, \citenamefont {Kim}, \citenamefont
  {Swagten},\ and\ \citenamefont {Koopmans}}]{Ummelen15}%
  \BibitemOpen
  \bibfield  {author} {\bibinfo {author} {\bibfnamefont {F.~C.}\ \bibnamefont
  {Ummelen}}, \bibinfo {author} {\bibfnamefont {D.-S.}\ \bibnamefont {Han}},
  \bibinfo {author} {\bibfnamefont {J.-S.}\ \bibnamefont {Kim}}, \bibinfo
  {author} {\bibfnamefont {H.~J.~M.}\ \bibnamefont {Swagten}}, \ and\ \bibinfo
  {author} {\bibfnamefont {B.}~\bibnamefont {Koopmans}},\ }\href@noop {}
  {\bibfield  {journal} {\bibinfo  {journal} {IEEE Trans. Magn.}\ }\textbf
  {\bibinfo {volume} {51}},\ \bibinfo {pages} {6000703} (\bibinfo {year}
  {2015})}\BibitemShut {NoStop}%
\bibitem [{\citenamefont {Emori}\ \emph {et~al.}(2013)\citenamefont {Emori},
  \citenamefont {Bauer}, \citenamefont {Ahn}, \citenamefont {Martinez},\ and\
  \citenamefont {Beach}}]{Emori13}%
  \BibitemOpen
  \bibfield  {author} {\bibinfo {author} {\bibfnamefont {S.}~\bibnamefont
  {Emori}}, \bibinfo {author} {\bibfnamefont {U.}~\bibnamefont {Bauer}},
  \bibinfo {author} {\bibfnamefont {S.-M.}\ \bibnamefont {Ahn}}, \bibinfo
  {author} {\bibfnamefont {E.}~\bibnamefont {Martinez}}, \ and\ \bibinfo
  {author} {\bibfnamefont {G.~S.~D.}\ \bibnamefont {Beach}},\ }\href@noop {}
  {\bibfield  {journal} {\bibinfo  {journal} {Nat. Mater.}\ }\textbf {\bibinfo
  {volume} {12}},\ \bibinfo {pages} {611} (\bibinfo {year} {2013})}\BibitemShut
  {NoStop}%
\bibitem [{\citenamefont {Thiaville}\ \emph {et~al.}(2012)\citenamefont
  {Thiaville}, \citenamefont {Rohart}, \citenamefont {Ju\'{e}}, \citenamefont
  {Cros},\ and\ \citenamefont {Fert}}]{Thiaville12}%
  \BibitemOpen
  \bibfield  {author} {\bibinfo {author} {\bibfnamefont {A.}~\bibnamefont
  {Thiaville}}, \bibinfo {author} {\bibfnamefont {S.}~\bibnamefont {Rohart}},
  \bibinfo {author} {\bibfnamefont {E.}~\bibnamefont {Ju\'{e}}}, \bibinfo
  {author} {\bibfnamefont {V.}~\bibnamefont {Cros}}, \ and\ \bibinfo {author}
  {\bibfnamefont {A.}~\bibnamefont {Fert}},\ }\href@noop {} {\bibfield
  {journal} {\bibinfo  {journal} {Europhys. Lett.}\ }\textbf {\bibinfo {volume}
  {100}},\ \bibinfo {pages} {57002} (\bibinfo {year} {2012})}\BibitemShut
  {NoStop}%
\bibitem [{\citenamefont {Sampaio}\ \emph {et~al.}(2013)\citenamefont
  {Sampaio}, \citenamefont {Cros}, \citenamefont {Rohart}, \citenamefont
  {Thiaville},\ and\ \citenamefont {Fert}}]{Sampaio13}%
  \BibitemOpen
  \bibfield  {author} {\bibinfo {author} {\bibfnamefont {J.}~\bibnamefont
  {Sampaio}}, \bibinfo {author} {\bibfnamefont {V.}~\bibnamefont {Cros}},
  \bibinfo {author} {\bibfnamefont {S.}~\bibnamefont {Rohart}}, \bibinfo
  {author} {\bibfnamefont {A.}~\bibnamefont {Thiaville}}, \ and\ \bibinfo
  {author} {\bibfnamefont {A.}~\bibnamefont {Fert}},\ }\href@noop {} {\bibfield
   {journal} {\bibinfo  {journal} {Nat. Nanotechnol.}\ }\textbf {\bibinfo
  {volume} {8}},\ \bibinfo {pages} {839} (\bibinfo {year} {2013})}\BibitemShut
  {NoStop}%
\bibitem [{\citenamefont {Jiang}\ \emph {et~al.}(2015)\citenamefont {Jiang},
  \citenamefont {Upadhyaya}, \citenamefont {Zhang}, \citenamefont {Yu},
  \citenamefont {Jungfleisch}, \citenamefont {Fradin}, \citenamefont {Pearson},
  \citenamefont {Tserkovnyak}, \citenamefont {Wang}, \citenamefont {Heinonen},
  \citenamefont {te~Velthuis},\ and\ \citenamefont {Hoffmann}}]{Jiang15}%
  \BibitemOpen
  \bibfield  {author} {\bibinfo {author} {\bibfnamefont {W.}~\bibnamefont
  {Jiang}}, \bibinfo {author} {\bibfnamefont {P.}~\bibnamefont {Upadhyaya}},
  \bibinfo {author} {\bibfnamefont {W.}~\bibnamefont {Zhang}}, \bibinfo
  {author} {\bibfnamefont {G.}~\bibnamefont {Yu}}, \bibinfo {author}
  {\bibfnamefont {M.}~\bibnamefont {Jungfleisch}}, \bibinfo {author}
  {\bibfnamefont {F.}~\bibnamefont {Fradin}}, \bibinfo {author} {\bibfnamefont
  {J.}~\bibnamefont {Pearson}}, \bibinfo {author} {\bibfnamefont
  {Y.}~\bibnamefont {Tserkovnyak}}, \bibinfo {author} {\bibfnamefont
  {K.}~\bibnamefont {Wang}}, \bibinfo {author} {\bibfnamefont {O.}~\bibnamefont
  {Heinonen}}, \bibinfo {author} {\bibfnamefont {S.}~\bibnamefont
  {te~Velthuis}}, \ and\ \bibinfo {author} {\bibfnamefont {A.}~\bibnamefont
  {Hoffmann}},\ }\href@noop {} {\bibfield  {journal} {\bibinfo  {journal}
  {Science}\ }\textbf {\bibinfo {volume} {349}},\ \bibinfo {pages} {283}
  (\bibinfo {year} {2015})}\BibitemShut {NoStop}%
\bibitem [{\citenamefont {Moreau-Luchaire}\ \emph {et~al.}(2016)\citenamefont
  {Moreau-Luchaire}, \citenamefont {Moutafis}, \citenamefont {Reyren},
  \citenamefont {Sampaio}, \citenamefont {Vaz}, \citenamefont {Van~Horne},
  \citenamefont {Bouzehouane}, \citenamefont {Garcia}, \citenamefont
  {Deranlot}, \citenamefont {Warnicke}, \citenamefont {Wohlh\"{u}ter},
  \citenamefont {George}, \citenamefont {Weigand}, \citenamefont {Raabe},
  \citenamefont {Cros},\ and\ \citenamefont {Fert}}]{Moreau-Luchaire16}%
  \BibitemOpen
  \bibfield  {author} {\bibinfo {author} {\bibfnamefont {C.}~\bibnamefont
  {Moreau-Luchaire}}, \bibinfo {author} {\bibfnamefont {C.}~\bibnamefont
  {Moutafis}}, \bibinfo {author} {\bibfnamefont {N.}~\bibnamefont {Reyren}},
  \bibinfo {author} {\bibfnamefont {J.}~\bibnamefont {Sampaio}}, \bibinfo
  {author} {\bibfnamefont {C.}~\bibnamefont {Vaz}}, \bibinfo {author}
  {\bibfnamefont {N.}~\bibnamefont {Van~Horne}}, \bibinfo {author}
  {\bibfnamefont {K.}~\bibnamefont {Bouzehouane}}, \bibinfo {author}
  {\bibfnamefont {K.}~\bibnamefont {Garcia}}, \bibinfo {author} {\bibfnamefont
  {C.}~\bibnamefont {Deranlot}}, \bibinfo {author} {\bibfnamefont
  {P.}~\bibnamefont {Warnicke}}, \bibinfo {author} {\bibfnamefont
  {P.}~\bibnamefont {Wohlh\"{u}ter}}, \bibinfo {author} {\bibfnamefont {J.-M.}\
  \bibnamefont {George}}, \bibinfo {author} {\bibfnamefont {M.}~\bibnamefont
  {Weigand}}, \bibinfo {author} {\bibfnamefont {J.}~\bibnamefont {Raabe}},
  \bibinfo {author} {\bibfnamefont {V.}~\bibnamefont {Cros}}, \ and\ \bibinfo
  {author} {\bibfnamefont {A.}~\bibnamefont {Fert}},\ }\href@noop {} {\bibfield
   {journal} {\bibinfo  {journal} {Nat. Nanotechnol.}\ }\textbf {\bibinfo
  {volume} {11}},\ \bibinfo {pages} {444} (\bibinfo {year} {2016})}\BibitemShut
  {NoStop}%
\bibitem [{\citenamefont {Woo}\ \emph {et~al.}(2016)\citenamefont {Woo},
  \citenamefont {Litzius}, \citenamefont {Kr\"{u}ger}, \citenamefont {Im},
  \citenamefont {Caretta}, \citenamefont {Richter}, \citenamefont {Mann},
  \citenamefont {Krone}, \citenamefont {Reeve}, \citenamefont {Weigand},
  \citenamefont {Agrawal}, \citenamefont {Lemesh}, \citenamefont {Mawass},
  \citenamefont {Fischer}, \citenamefont {Kl\"{a}ui},\ and\ \citenamefont
  {Beach}}]{Woo16}%
  \BibitemOpen
  \bibfield  {author} {\bibinfo {author} {\bibfnamefont {S.}~\bibnamefont
  {Woo}}, \bibinfo {author} {\bibfnamefont {K.}~\bibnamefont {Litzius}},
  \bibinfo {author} {\bibfnamefont {B.}~\bibnamefont {Kr\"{u}ger}}, \bibinfo
  {author} {\bibfnamefont {M.-Y.}\ \bibnamefont {Im}}, \bibinfo {author}
  {\bibfnamefont {L.}~\bibnamefont {Caretta}}, \bibinfo {author} {\bibfnamefont
  {K.}~\bibnamefont {Richter}}, \bibinfo {author} {\bibfnamefont
  {M.}~\bibnamefont {Mann}}, \bibinfo {author} {\bibfnamefont {A.}~\bibnamefont
  {Krone}}, \bibinfo {author} {\bibfnamefont {R.}~\bibnamefont {Reeve}},
  \bibinfo {author} {\bibfnamefont {M.}~\bibnamefont {Weigand}}, \bibinfo
  {author} {\bibfnamefont {P.}~\bibnamefont {Agrawal}}, \bibinfo {author}
  {\bibfnamefont {I.}~\bibnamefont {Lemesh}}, \bibinfo {author} {\bibfnamefont
  {M.-A.}\ \bibnamefont {Mawass}}, \bibinfo {author} {\bibfnamefont
  {P.}~\bibnamefont {Fischer}}, \bibinfo {author} {\bibfnamefont
  {M.}~\bibnamefont {Kl\"{a}ui}}, \ and\ \bibinfo {author} {\bibfnamefont
  {G.}~\bibnamefont {Beach}},\ }\href@noop {} {\bibfield  {journal} {\bibinfo
  {journal} {Nat. Mater.}\ }\textbf {\bibinfo {volume} {15}},\ \bibinfo {pages}
  {501} (\bibinfo {year} {2016})}\BibitemShut {NoStop}%
\bibitem [{\citenamefont {Boulle}\ \emph {et~al.}(2016)\citenamefont {Boulle},
  \citenamefont {Vogel}, \citenamefont {Yang}, \citenamefont {Pizzini},
  \citenamefont {de~Souza~Chaves}, \citenamefont {Locatelli}, \citenamefont
  {Mentes}, \citenamefont {Sala}, \citenamefont {Buda-Prejbeanu}, \citenamefont
  {Klein}, \citenamefont {Belmeguenai}, \citenamefont {Roussign\'{e}},
  \citenamefont {Stashkevich}, \citenamefont {Ch\'{e}rif}, \citenamefont
  {Aballe}, \citenamefont {Foerster}, \citenamefont {Chshiev}, \citenamefont
  {Auffret}, \citenamefont {Miron},\ and\ \citenamefont {Gaudin}}]{Boulle16}%
  \BibitemOpen
  \bibfield  {author} {\bibinfo {author} {\bibfnamefont {O.}~\bibnamefont
  {Boulle}}, \bibinfo {author} {\bibfnamefont {J.}~\bibnamefont {Vogel}},
  \bibinfo {author} {\bibfnamefont {H.}~\bibnamefont {Yang}}, \bibinfo {author}
  {\bibfnamefont {S.}~\bibnamefont {Pizzini}}, \bibinfo {author} {\bibfnamefont
  {D.}~\bibnamefont {de~Souza~Chaves}}, \bibinfo {author} {\bibfnamefont
  {A.}~\bibnamefont {Locatelli}}, \bibinfo {author} {\bibfnamefont
  {T.}~\bibnamefont {Mentes}}, \bibinfo {author} {\bibfnamefont
  {A.}~\bibnamefont {Sala}}, \bibinfo {author} {\bibfnamefont {L.}~\bibnamefont
  {Buda-Prejbeanu}}, \bibinfo {author} {\bibfnamefont {O.}~\bibnamefont
  {Klein}}, \bibinfo {author} {\bibfnamefont {M.}~\bibnamefont {Belmeguenai}},
  \bibinfo {author} {\bibfnamefont {Y.}~\bibnamefont {Roussign\'{e}}}, \bibinfo
  {author} {\bibfnamefont {A.}~\bibnamefont {Stashkevich}}, \bibinfo {author}
  {\bibfnamefont {S.}~\bibnamefont {Ch\'{e}rif}}, \bibinfo {author}
  {\bibfnamefont {L.}~\bibnamefont {Aballe}}, \bibinfo {author} {\bibfnamefont
  {M.}~\bibnamefont {Foerster}}, \bibinfo {author} {\bibfnamefont
  {M.}~\bibnamefont {Chshiev}}, \bibinfo {author} {\bibfnamefont
  {S.}~\bibnamefont {Auffret}}, \bibinfo {author} {\bibfnamefont
  {I.}~\bibnamefont {Miron}}, \ and\ \bibinfo {author} {\bibfnamefont
  {G.}~\bibnamefont {Gaudin}},\ }\href@noop {} {\bibfield  {journal} {\bibinfo
  {journal} {Nat. Nanotechnol.}\ }\textbf {\bibinfo {volume} {11}},\ \bibinfo
  {pages} {449} (\bibinfo {year} {2016})}\BibitemShut {NoStop}%
\bibitem [{\citenamefont {Kang}\ \emph {et~al.}(2016)\citenamefont {Kang},
  \citenamefont {Huang}, \citenamefont {Zheng}, \citenamefont {Lv},
  \citenamefont {Lei}, \citenamefont {Zhang}, \citenamefont {Zhang},
  \citenamefont {Zhou},\ and\ \citenamefont {Zhao}}]{Kang16}%
  \BibitemOpen
  \bibfield  {author} {\bibinfo {author} {\bibfnamefont {W.}~\bibnamefont
  {Kang}}, \bibinfo {author} {\bibfnamefont {Y.}~\bibnamefont {Huang}},
  \bibinfo {author} {\bibfnamefont {C.}~\bibnamefont {Zheng}}, \bibinfo
  {author} {\bibfnamefont {W.}~\bibnamefont {Lv}}, \bibinfo {author}
  {\bibfnamefont {N.}~\bibnamefont {Lei}}, \bibinfo {author} {\bibfnamefont
  {Y.}~\bibnamefont {Zhang}}, \bibinfo {author} {\bibfnamefont
  {X.}~\bibnamefont {Zhang}}, \bibinfo {author} {\bibfnamefont
  {Y.}~\bibnamefont {Zhou}}, \ and\ \bibinfo {author} {\bibfnamefont
  {W.}~\bibnamefont {Zhao}},\ }\href@noop {} {\bibfield  {journal} {\bibinfo
  {journal} {Sci. Rep.}\ }\textbf {\bibinfo {volume} {6}},\ \bibinfo {pages}
  {23164} (\bibinfo {year} {2016})}\BibitemShut {NoStop}%
\bibitem [{\citenamefont {Parkin}\ \emph {et~al.}(2008)\citenamefont {Parkin},
  \citenamefont {Hayashi},\ and\ \citenamefont {Thomas}}]{Parkin08}%
  \BibitemOpen
  \bibfield  {author} {\bibinfo {author} {\bibfnamefont {S.~S.~P.}\
  \bibnamefont {Parkin}}, \bibinfo {author} {\bibfnamefont {M.}~\bibnamefont
  {Hayashi}}, \ and\ \bibinfo {author} {\bibfnamefont {L.}~\bibnamefont
  {Thomas}},\ }\href@noop {} {\bibfield  {journal} {\bibinfo  {journal}
  {Science}\ }\textbf {\bibinfo {volume} {320}},\ \bibinfo {pages} {190}
  (\bibinfo {year} {2008})}\BibitemShut {NoStop}%
\bibitem [{\citenamefont {Kang}\ \emph {et~al.}(2015)\citenamefont {Kang},
  \citenamefont {Huang}, \citenamefont {Zheng}, \citenamefont {Lv},
  \citenamefont {Lei}, \citenamefont {Zhang}, \citenamefont {Zhang},
  \citenamefont {Zhou},\ and\ \citenamefont {Zhao}}]{Zhang15}%
  \BibitemOpen
  \bibfield  {author} {\bibinfo {author} {\bibfnamefont {W.}~\bibnamefont
  {Kang}}, \bibinfo {author} {\bibfnamefont {Y.}~\bibnamefont {Huang}},
  \bibinfo {author} {\bibfnamefont {C.}~\bibnamefont {Zheng}}, \bibinfo
  {author} {\bibfnamefont {W.}~\bibnamefont {Lv}}, \bibinfo {author}
  {\bibfnamefont {N.}~\bibnamefont {Lei}}, \bibinfo {author} {\bibfnamefont
  {Y.}~\bibnamefont {Zhang}}, \bibinfo {author} {\bibfnamefont
  {X.}~\bibnamefont {Zhang}}, \bibinfo {author} {\bibfnamefont
  {Y.}~\bibnamefont {Zhou}}, \ and\ \bibinfo {author} {\bibfnamefont
  {W.}~\bibnamefont {Zhao}},\ }\href@noop {} {\bibfield  {journal} {\bibinfo
  {journal} {Sci. Rep.}\ }\textbf {\bibinfo {volume} {6}},\ \bibinfo {pages}
  {9400} (\bibinfo {year} {2015})}\BibitemShut {NoStop}%
\bibitem [{\citenamefont {Allwood}\ \emph {et~al.}(2005)\citenamefont
  {Allwood}, \citenamefont {Xiong}, \citenamefont {Faulkner}, \citenamefont
  {Atkinson}, \citenamefont {Petit},\ and\ \citenamefont
  {Cowburn}}]{Allwood05}%
  \BibitemOpen
  \bibfield  {author} {\bibinfo {author} {\bibfnamefont {D.~A.}\ \bibnamefont
  {Allwood}}, \bibinfo {author} {\bibfnamefont {G.}~\bibnamefont {Xiong}},
  \bibinfo {author} {\bibfnamefont {C.~C.}\ \bibnamefont {Faulkner}}, \bibinfo
  {author} {\bibfnamefont {D.}~\bibnamefont {Atkinson}}, \bibinfo {author}
  {\bibfnamefont {D.}~\bibnamefont {Petit}}, \ and\ \bibinfo {author}
  {\bibfnamefont {R.~P.}\ \bibnamefont {Cowburn}},\ }\href@noop {} {\bibfield
  {journal} {\bibinfo  {journal} {Science}\ }\textbf {\bibinfo {volume}
  {309}},\ \bibinfo {pages} {1688} (\bibinfo {year} {2005})}\BibitemShut
  {NoStop}%
\bibitem [{\citenamefont {Fert}(1990)}]{Fert90}%
  \BibitemOpen
  \bibfield  {author} {\bibinfo {author} {\bibfnamefont {A.~R.}\ \bibnamefont
  {Fert}},\ }\href@noop {} {\bibfield  {journal} {\bibinfo  {journal} {Mater.
  Sci. Forum}\ }\textbf {\bibinfo {volume} {59-60}},\ \bibinfo {pages} {439}
  (\bibinfo {year} {1990})}\BibitemShut {NoStop}%
\bibitem [{\citenamefont {Moon}\ \emph {et~al.}(2015)\citenamefont {Moon},
  \citenamefont {Kim}, \citenamefont {Yoo}, \citenamefont {Je}, \citenamefont
  {Chun}, \citenamefont {Kim}, \citenamefont {Min}, \citenamefont {Hwang},\
  and\ \citenamefont {Choe}}]{Moon15}%
  \BibitemOpen
  \bibfield  {author} {\bibinfo {author} {\bibfnamefont {K.-W.}\ \bibnamefont
  {Moon}}, \bibinfo {author} {\bibfnamefont {D.-H.}\ \bibnamefont {Kim}},
  \bibinfo {author} {\bibfnamefont {S.-C.}\ \bibnamefont {Yoo}}, \bibinfo
  {author} {\bibfnamefont {S.-G.}\ \bibnamefont {Je}}, \bibinfo {author}
  {\bibfnamefont {B.~S.}\ \bibnamefont {Chun}}, \bibinfo {author}
  {\bibfnamefont {W.}~\bibnamefont {Kim}}, \bibinfo {author} {\bibfnamefont
  {B.-C.}\ \bibnamefont {Min}}, \bibinfo {author} {\bibfnamefont
  {C.}~\bibnamefont {Hwang}}, \ and\ \bibinfo {author} {\bibfnamefont {S.-B.}\
  \bibnamefont {Choe}},\ }\href@noop {} {\bibfield  {journal} {\bibinfo
  {journal} {Sci. Rep.}\ }\textbf {\bibinfo {volume} {5}},\ \bibinfo {pages}
  {9166} (\bibinfo {year} {2015})}\BibitemShut {NoStop}%
\bibitem [{\citenamefont {Petit}\ \emph {et~al.}(2015)\citenamefont {Petit},
  \citenamefont {Seem}, \citenamefont {Tillette}, \citenamefont {Mansell},\
  and\ \citenamefont {Cowburn}}]{Petit15}%
  \BibitemOpen
  \bibfield  {author} {\bibinfo {author} {\bibfnamefont {D.}~\bibnamefont
  {Petit}}, \bibinfo {author} {\bibfnamefont {P.~R.}\ \bibnamefont {Seem}},
  \bibinfo {author} {\bibfnamefont {M.}~\bibnamefont {Tillette}}, \bibinfo
  {author} {\bibfnamefont {R.}~\bibnamefont {Mansell}}, \ and\ \bibinfo
  {author} {\bibfnamefont {R.~P.}\ \bibnamefont {Cowburn}},\ }\href@noop {}
  {\bibfield  {journal} {\bibinfo  {journal} {Appl. Phys. Lett.}\ }\textbf
  {\bibinfo {volume} {106}},\ \bibinfo {pages} {022402} (\bibinfo {year}
  {2015})}\BibitemShut {NoStop}%
\bibitem [{\citenamefont {Mihai}\ \emph {et~al.}(2013)\citenamefont {Mihai},
  \citenamefont {Whiteside}, \citenamefont {Canwell}, \citenamefont {Marrows},
  \citenamefont {Benitez}, \citenamefont {McGrouther}, \citenamefont {McVitie},
  \citenamefont {McFadzean},\ and\ \citenamefont {Moore}}]{Mihai13}%
  \BibitemOpen
  \bibfield  {author} {\bibinfo {author} {\bibfnamefont {A.~P.}\ \bibnamefont
  {Mihai}}, \bibinfo {author} {\bibfnamefont {A.~L.}\ \bibnamefont
  {Whiteside}}, \bibinfo {author} {\bibfnamefont {E.~J.}\ \bibnamefont
  {Canwell}}, \bibinfo {author} {\bibfnamefont {C.~H.}\ \bibnamefont
  {Marrows}}, \bibinfo {author} {\bibfnamefont {M.~J.}\ \bibnamefont
  {Benitez}}, \bibinfo {author} {\bibfnamefont {D.}~\bibnamefont {McGrouther}},
  \bibinfo {author} {\bibfnamefont {S.}~\bibnamefont {McVitie}}, \bibinfo
  {author} {\bibfnamefont {S.}~\bibnamefont {McFadzean}}, \ and\ \bibinfo
  {author} {\bibfnamefont {T.~A.}\ \bibnamefont {Moore}},\ }\href@noop {}
  {\bibfield  {journal} {\bibinfo  {journal} {Appl. Phys. Lett.}\ }\textbf
  {\bibinfo {volume} {103}},\ \bibinfo {pages} {262401} (\bibinfo {year}
  {2013})}\BibitemShut {NoStop}%
\bibitem [{\citenamefont {Sinha}\ \emph {et~al.}(1988)\citenamefont {Sinha},
  \citenamefont {Sirota}, \citenamefont {Garoff},\ and\ \citenamefont
  {Stanley}}]{Sinha88}%
  \BibitemOpen
  \bibfield  {author} {\bibinfo {author} {\bibfnamefont {S.~K.}\ \bibnamefont
  {Sinha}}, \bibinfo {author} {\bibfnamefont {E.~B.}\ \bibnamefont {Sirota}},
  \bibinfo {author} {\bibfnamefont {S.}~\bibnamefont {Garoff}}, \ and\ \bibinfo
  {author} {\bibfnamefont {H.~B.}\ \bibnamefont {Stanley}},\ }\href@noop {}
  {\bibfield  {journal} {\bibinfo  {journal} {Phys. Rev. B}\ }\textbf {\bibinfo
  {volume} {38}},\ \bibinfo {pages} {2297} (\bibinfo {year}
  {1988})}\BibitemShut {NoStop}%
\bibitem [{\citenamefont {Savage}\ \emph {et~al.}(1991)\citenamefont {Savage},
  \citenamefont {Kleiner}, \citenamefont {Schimke}, \citenamefont {Phang},
  \citenamefont {Jankowski}, \citenamefont {Jacobs}, \citenamefont {Kariotis},\
  and\ \citenamefont {Lagally}}]{Savage91}%
  \BibitemOpen
  \bibfield  {author} {\bibinfo {author} {\bibfnamefont {D.~E.}\ \bibnamefont
  {Savage}}, \bibinfo {author} {\bibfnamefont {J.}~\bibnamefont {Kleiner}},
  \bibinfo {author} {\bibfnamefont {N.}~\bibnamefont {Schimke}}, \bibinfo
  {author} {\bibfnamefont {Y.-H.}\ \bibnamefont {Phang}}, \bibinfo {author}
  {\bibfnamefont {T.}~\bibnamefont {Jankowski}}, \bibinfo {author}
  {\bibfnamefont {J.}~\bibnamefont {Jacobs}}, \bibinfo {author} {\bibfnamefont
  {R.}~\bibnamefont {Kariotis}}, \ and\ \bibinfo {author} {\bibfnamefont
  {M.~G.}\ \bibnamefont {Lagally}},\ }\href@noop {} {\bibfield  {journal}
  {\bibinfo  {journal} {J. Appl. Phys.}\ }\textbf {\bibinfo {volume} {69}},\
  \bibinfo {pages} {1411} (\bibinfo {year} {1991})}\BibitemShut {NoStop}%
\bibitem [{\citenamefont {Wormington}\ \emph {et~al.}(1999)\citenamefont
  {Wormington}, \citenamefont {Panaccione}, \citenamefont {Matney},\ and\
  \citenamefont {Bowen}}]{Wormington99}%
  \BibitemOpen
  \bibfield  {author} {\bibinfo {author} {\bibfnamefont {M.}~\bibnamefont
  {Wormington}}, \bibinfo {author} {\bibfnamefont {C.}~\bibnamefont
  {Panaccione}}, \bibinfo {author} {\bibfnamefont {K.~M.}\ \bibnamefont
  {Matney}}, \ and\ \bibinfo {author} {\bibfnamefont {D.~K.}\ \bibnamefont
  {Bowen}},\ }\href@noop {} {\bibfield  {journal} {\bibinfo  {journal} {Phil.
  Trans. R. Soc. A}\ }\textbf {\bibinfo {volume} {357}},\ \bibinfo {pages}
  {2827} (\bibinfo {year} {1999})}\BibitemShut {NoStop}%
\bibitem [{\citenamefont {Bolzoni}\ \emph {et~al.}(1984)\citenamefont
  {Bolzoni}, \citenamefont {Leccabue}, \citenamefont {Panizzieri},\ and\
  \citenamefont {Pareti}}]{Bolzoni84}%
  \BibitemOpen
  \bibfield  {author} {\bibinfo {author} {\bibfnamefont {F.}~\bibnamefont
  {Bolzoni}}, \bibinfo {author} {\bibfnamefont {F.}~\bibnamefont {Leccabue}},
  \bibinfo {author} {\bibfnamefont {R.}~\bibnamefont {Panizzieri}}, \ and\
  \bibinfo {author} {\bibfnamefont {L.}~\bibnamefont {Pareti}},\ }\href@noop {}
  {\bibfield  {journal} {\bibinfo  {journal} {IEEE Trans. Magn.}\ }\textbf
  {\bibinfo {volume} {20}},\ \bibinfo {pages} {1625} (\bibinfo {year}
  {1984})}\BibitemShut {NoStop}%
\bibitem [{\citenamefont {Moon}\ \emph {et~al.}(2009)\citenamefont {Moon},
  \citenamefont {Lee}, \citenamefont {Choe},\ and\ \citenamefont
  {Shin}}]{Moon09}%
  \BibitemOpen
  \bibfield  {author} {\bibinfo {author} {\bibfnamefont {K.-W.}\ \bibnamefont
  {Moon}}, \bibinfo {author} {\bibfnamefont {J.-C.}\ \bibnamefont {Lee}},
  \bibinfo {author} {\bibfnamefont {S.-B.}\ \bibnamefont {Choe}}, \ and\
  \bibinfo {author} {\bibfnamefont {K.-H.}\ \bibnamefont {Shin}},\ }\href@noop
  {} {\bibfield  {journal} {\bibinfo  {journal} {Rev. Sci. Instrum.}\ }\textbf
  {\bibinfo {volume} {80}},\ \bibinfo {pages} {113904} (\bibinfo {year}
  {2009})}\BibitemShut {NoStop}%
\bibitem [{\citenamefont {Ju\'{e}}\ \emph {et~al.}(2016)\citenamefont
  {Ju\'{e}}, \citenamefont {Safeer}, \citenamefont {Drouard}, \citenamefont
  {Lopez}, \citenamefont {Balint}, \citenamefont {Buda-Prejbeanu},
  \citenamefont {Boulle}, \citenamefont {Auffret}, \citenamefont {Schuhl},
  \citenamefont {Manchon}, \citenamefont {Miron},\ and\ \citenamefont
  {Gaudin}}]{Jue15}%
  \BibitemOpen
  \bibfield  {author} {\bibinfo {author} {\bibfnamefont {E.}~\bibnamefont
  {Ju\'{e}}}, \bibinfo {author} {\bibfnamefont {C.~K.}\ \bibnamefont {Safeer}},
  \bibinfo {author} {\bibfnamefont {M.}~\bibnamefont {Drouard}}, \bibinfo
  {author} {\bibfnamefont {A.}~\bibnamefont {Lopez}}, \bibinfo {author}
  {\bibfnamefont {P.}~\bibnamefont {Balint}}, \bibinfo {author} {\bibfnamefont
  {L.}~\bibnamefont {Buda-Prejbeanu}}, \bibinfo {author} {\bibfnamefont
  {O.}~\bibnamefont {Boulle}}, \bibinfo {author} {\bibfnamefont
  {S.}~\bibnamefont {Auffret}}, \bibinfo {author} {\bibfnamefont
  {A.}~\bibnamefont {Schuhl}}, \bibinfo {author} {\bibfnamefont
  {A.}~\bibnamefont {Manchon}}, \bibinfo {author} {\bibfnamefont {I.~M.}\
  \bibnamefont {Miron}}, \ and\ \bibinfo {author} {\bibfnamefont
  {G.}~\bibnamefont {Gaudin}},\ }\href@noop {} {\bibfield  {journal} {\bibinfo
  {journal} {Nat. Mater.}\ }\textbf {\bibinfo {volume} {15}},\ \bibinfo {pages}
  {272} (\bibinfo {year} {2016})}\BibitemShut {NoStop}%
\bibitem [{\citenamefont {Lau}\ \emph {et~al.}(2016)\citenamefont {Lau},
  \citenamefont {Sundar}, \citenamefont {Zhu},\ and\ \citenamefont
  {Sokalski}}]{Lau16}%
  \BibitemOpen
  \bibfield  {author} {\bibinfo {author} {\bibfnamefont {D.}~\bibnamefont
  {Lau}}, \bibinfo {author} {\bibfnamefont {V.}~\bibnamefont {Sundar}},
  \bibinfo {author} {\bibfnamefont {J.-G.}\ \bibnamefont {Zhu}}, \ and\
  \bibinfo {author} {\bibfnamefont {V.}~\bibnamefont {Sokalski}},\ }\href@noop
  {} {\bibfield  {journal} {\bibinfo  {journal} {Phys. Rev. B}\ }\textbf
  {\bibinfo {volume} {94}},\ \bibinfo {pages} {060401(R)} (\bibinfo {year}
  {2016})}\BibitemShut {NoStop}%
\bibitem [{\citenamefont {Cayssol}\ \emph {et~al.}(2004)\citenamefont
  {Cayssol}, \citenamefont {Ravelosona}, \citenamefont {Chappert},
  \citenamefont {Ferr\'{e}},\ and\ \citenamefont {Jamet}}]{Cayssol04}%
  \BibitemOpen
  \bibfield  {author} {\bibinfo {author} {\bibfnamefont {F.}~\bibnamefont
  {Cayssol}}, \bibinfo {author} {\bibfnamefont {D.}~\bibnamefont {Ravelosona}},
  \bibinfo {author} {\bibfnamefont {C.}~\bibnamefont {Chappert}}, \bibinfo
  {author} {\bibfnamefont {J.}~\bibnamefont {Ferr\'{e}}}, \ and\ \bibinfo
  {author} {\bibfnamefont {J.~P.}\ \bibnamefont {Jamet}},\ }\href@noop {}
  {\bibfield  {journal} {\bibinfo  {journal} {Phys. Rev. Lett.}\ }\textbf
  {\bibinfo {volume} {92}},\ \bibinfo {pages} {107202} (\bibinfo {year}
  {2004})}\BibitemShut {NoStop}%
\bibitem [{\citenamefont {Metaxas}\ \emph {et~al.}(2007)\citenamefont
  {Metaxas}, \citenamefont {Jamet}, \citenamefont {Mougin}, \citenamefont
  {Cormier}, \citenamefont {Ferr\'{e}}, \citenamefont {Baltz}, \citenamefont
  {Rodmacq}, \citenamefont {Dieny},\ and\ \citenamefont {Stamps}}]{Metaxas07}%
  \BibitemOpen
  \bibfield  {author} {\bibinfo {author} {\bibfnamefont {P.~J.}\ \bibnamefont
  {Metaxas}}, \bibinfo {author} {\bibfnamefont {J.~P.}\ \bibnamefont {Jamet}},
  \bibinfo {author} {\bibfnamefont {A.}~\bibnamefont {Mougin}}, \bibinfo
  {author} {\bibfnamefont {M.}~\bibnamefont {Cormier}}, \bibinfo {author}
  {\bibfnamefont {J.}~\bibnamefont {Ferr\'{e}}}, \bibinfo {author}
  {\bibfnamefont {V.}~\bibnamefont {Baltz}}, \bibinfo {author} {\bibfnamefont
  {B.}~\bibnamefont {Rodmacq}}, \bibinfo {author} {\bibfnamefont
  {B.}~\bibnamefont {Dieny}}, \ and\ \bibinfo {author} {\bibfnamefont {R.~L.}\
  \bibnamefont {Stamps}},\ }\href@noop {} {\bibfield  {journal} {\bibinfo
  {journal} {Phys. Rev. Lett.}\ }\textbf {\bibinfo {volume} {99}},\ \bibinfo
  {pages} {217208} (\bibinfo {year} {2007})}\BibitemShut {NoStop}%
\bibitem [{\citenamefont {Franken}\ \emph {et~al.}(2014)\citenamefont
  {Franken}, \citenamefont {Herps}, \citenamefont {Swagten},\ and\
  \citenamefont {Koopmans}}]{Franken14}%
  \BibitemOpen
  \bibfield  {author} {\bibinfo {author} {\bibfnamefont {J.~H.}\ \bibnamefont
  {Franken}}, \bibinfo {author} {\bibfnamefont {M.}~\bibnamefont {Herps}},
  \bibinfo {author} {\bibfnamefont {H.~J.~M.}\ \bibnamefont {Swagten}}, \ and\
  \bibinfo {author} {\bibfnamefont {B.}~\bibnamefont {Koopmans}},\ }\href@noop
  {} {\bibfield  {journal} {\bibinfo  {journal} {Sci. Rep.}\ }\textbf {\bibinfo
  {volume} {4}},\ \bibinfo {pages} {5248} (\bibinfo {year} {2014})}\BibitemShut
  {NoStop}%
\end{thebibliography}%

\end{document}